\documentclass[a4paper,10pt,dvipsnames]{article}
\usepackage[utf8]{inputenc}
\usepackage[english]{babel}
\usepackage{amsthm}
\usepackage{amsfonts}
\usepackage{amssymb}	
\usepackage{amsmath}
\allowdisplaybreaks
\usepackage{mathtools}
\usepackage{slashed}
\usepackage{enumerate}
\usepackage{graphicx}
\usepackage{systeme}
\usepackage{dsfont}
\usepackage{relsize}
\usepackage[margin=0.90in]{geometry}
\usepackage{float}
\usepackage{tikz}
\usepackage[labelformat=simple]{subcaption}

\usepackage{graphicx}
\usepackage{bmpsize}
\usepackage{epstopdf}
\usepackage{bbm}
\usepackage{etoolbox}
\usepackage{xcolor}
\usepackage{titling}
\usepackage{authblk}

\usepackage{blindtext,graphicx}
\usepackage[absolute]{textpos}
\usepackage{physics}
\usepackage[hang,flushmargin]{footmisc}

\usepackage{xfrac}
\usepackage{nicefrac}
\usepackage{esvect}
\usepackage{cases}
\usepackage{empheq}
\usepackage{stmaryrd}
\usepackage{cancel}
\usepackage[linguistics]{forest}
\usepackage{url}
\usepackage[font=small]{caption}
\usepackage{array}
\usepackage[giveninits=true,sortcites=true,date=year,maxbibnames=99,doi=false,isbn=false,url=false,eprint=false]{biblatex}
\renewbibmacro{in:}{}
\DeclareFieldFormat{pages}{#1}
\AtEveryBibitem{%
  \clearlist{language}%
}
\usepackage{csquotes}

\usepackage{setspace}
\makeatletter
\newcommand{\thickhline}{%
    \noalign {\ifnum 0=`}\fi \hrule height 1pt
    \futurelet \reserved@a \@xhline
}
\newcolumntype{"}{@{\hskip\tabcolsep\vrule width 1pt\hskip\tabcolsep}}
\makeatother
\theoremstyle{definition}

\makeatletter
\newcommand{\pushright}[1]{\ifmeasuring@#1\else\omit\hfill$\displaystyle#1$\fi\ignorespaces}
\newcommand{\pushleft}[1]{\ifmeasuring@#1\else\omit$\displaystyle#1$\hfill\fi\ignorespaces}
\makeatother
\title{
Metamaterial shields for inner protection and outer tuning through a relaxed micromorphic approach
}
\author{
	Gianluca Rizzi\thanks{Faculty of Architecture and Civil Engineering, TU Dortmund, August-Schmidt-Str. 8, 44227 Dortmund, Germany},
	\quad
	Patrizio Neff\thanks{Head of Chair for Nonlinear Analysis and Modelling, Fakultät für Mathematik, Universität Duisburg-Essen, \\ \indent \,\,\,\, Thea-Leymann-Straße 9, 45127 Essen, Germany},
	\quad and \quad
	Angela Madeo\thanks{Head of Chair of Continuum Mechanics, Faculty of Architecture and Civil Engineering, TU Dortmund, \\ \indent \,\,\,\, August-Schmidt-Str. 8, 44227 Dortmund, Germany}
	}

\thanksmarkseries{arabic}
\date{\today}
\begin{document}
\maketitle
\begin{abstract}
In this paper, a coherent boundary value problem to model metamaterials’ behavior based on the relaxed micromorphic model is established.
This boundary value problem includes well-posed boundary conditions, thus disclosing the possibility of exploring the scattering patterns of finite-size metamaterials’ specimens.
Thanks to the simplified model’s structure (few frequency- and angle-independent parameters), we are able to unveil the scattering metamaterial’s response for a wide range of frequencies and angles of propagation of the incident wave.
These results are an important stepping stone towards the conception of more complex large-scale meta-structures that can control elastic waves and recover energy.
\end{abstract}

\section{Introduction}
\label{sec:intro}
For both bio-inspired \cite{miniaci2016spider} or regular metamaterials \cite{brun2012vortex,la2017conception}, a useful property that is often sought (by intuition or by a reverse design of the unit cell \cite{ragonese2021prediction}) is the ability to prevent the propagation of elastic waves.
In this way, the metamaterial can act as a shield if placed around an object that is wanted to be isolated from the external environment.
In the case of a periodic metamaterial, the ranges of frequencies for which no wave propagate (i.e. the band-gap intervals) can be inferred from the dispersion diagram of the unit cell.
The band-gap width can be tuned passively by acting on the geometry of the unit cell and on the materials constituting it \cite{rizzi_exploring_2021,rizzi2020towards,rizzi2021boundary}, by introducing an elastic prestress \cite{barnwell2016antiplane,bordiga2019prestress}, or actively by modifying these properties thanks to external solicitations in real time like in piezoelectric materials \cite{bacigalupo2020design}.

Since the last decade, mechanical metamaterials are at the center of cutting-edge research efforts to design devices interacting with elastic waves, such as shields \cite{colombi2016seismic,du2017elastic,miniaci2016large,achaoui2016seismic}, cloaks  \cite{buckmann_mechanical_2015,misseroni_cymatics_2016,zhou_elastic_2008,zhang_asymmetric_2020}, and lenses \cite{fang2006ultrasonic,dubois2013flat,jin2016gradient}.
Indeed, metamaterials’ heterogeneous microstructure can be today architectured to constructively interact with elastic waves, thus enabling the emergence of unorthodox responses at the macroscopic scale, including band-gaps, cloaking, focusing, channeling, negative refraction, and many more \cite{celli_bandgap_2019,bilal_architected_2018,liu_locally_2000,wang_harnessing_2014,buckmann_mechanical_2015,misseroni_cymatics_2016,zhou_elastic_2008,zhang_asymmetric_2020,cummer_controlling_2016,guenneau_acoustic_2007,kaina_slow_2017,tallarico_edge_2017,willis_negative_2016}.
One important application is that of exploiting metamaterials’ band-gaps to create elastic shielding devices protecting objects from elastic waves \cite{miniaci2016spider,brun2012vortex,la2017conception,ragonese2021prediction}.
Depending on the devices’ operating frequencies, these shields can act for seismic \cite{guenneau_acoustic_2007}, acoustic \cite{rizzi_exploring_2021,la2017conception,nolde_high_2011,sridhar_general_2018,sridhar_frequency_2020} or ultrasonic \cite{aivaliotis_frequency-_2020,rizzi_exploring_2021,rizzi2020towards,rizzi2021boundary,craster_high_frequency_2010} protection of objects that need to be isolated from the external environment.

In the light of the previous remarks, it is established that metamaterials can be used for protection purposes.
Yet, no effort is made to explore the energy which is reflected by the shield and that affects its surroundings.
It is of primary importance to master such reflection phenomena to make metamaterial shields suitable for practical purposes.
The effect of elastic waves in the surroundings of the shield is usually not explored due to the lack of simplified homogenized models allowing the simulation of large meta-structures with reasonable computational costs.
We show in this paper that, once such a model is established, suitable boundary value problems can be set up allowing an eased simulation of both the inner and outer domain.
In fact, even if dynamic homogenization methods have provided important advancements in the modeling of infinite-size metamaterials \cite{pham_transient_2013,sridhar_homogenization_2016,sridhar_general_2018,sridhar_frequency_2020,craster_high_frequency_2010,willis_effective_2011,willis_variational_1981,nassar_willis_2015}, little is known about handling scattering problems at the boundaries of finite-size metamaterials.
This lack of knowledge is mainly due to the difficulties arising in establishing pertinent macroscopic boundary conditions formally upscaled from microscopic ones.
To overcome this problem, we propose here to study the scattering properties of the considered shielding device via the relaxed micromorphic model, introduced by the authors \cite{neff_identification_2020,neff2014unifying,dagostino_effective_2020} and recently equipped with a set of well-posed, physically coherent boundary conditions \cite{rizzi2021boundary}.
We prove that the relaxed micromorphic model shows excellent agreement with the scattering profiles obtained via full microstructured simulations when a direct comparison is possible due to the small specimen’s size.
This agreement is established for a wide frequency range (from zero to beyond the first band-gap) and for all angles of propagation of the incident wave.
These results are important since, to the authors’ knowledge, no other homogenized model is able to predict finite-size metamaterials’response for all frequencies and angles of incidence with a restricted number of frequency- and angle-independent material parameters.
The versatility of our model allows us to treat otherwise impossible problems such as the exploration of shields of increasing size.
We eventually show that the pattern of reflected energy changes while increasing the size of the shield itself.

The paper is organized as follows.
In section \ref{sec:unit_cell_const_law} we present the unit cell studied, the constitutive law for the relaxed micromorphic  model and the elastic and the micro-inertia parameters issued from a fitting procedure done on the dispersion curves.
In section \ref{sec:comparison} we set up a time harmonic boundary value problem for a finite-size structure embedded in an infinitely large Cauchy medium and we present the comparison between the response of the real structure and the one made up of the equivalent relaxed micromorphic material for waves with three different frequencies and two different directions of propagation.
We then show meta-structures of increasing size that can be uniquely explored through the relaxed micromorphic model.
In section \ref{sec:conclusion} we draw the conclusions and we outline some perspectives.

\section{Modelling the mechanical response of an anisotropic metamaterial}
\label{sec:unit_cell_const_law}
Because of their inhomogeneous microstructures, mechanical metamaterials often show unorthodox elastic properties  in terms of wave propagation
\cite{celli_bandgap_2019,bilal_architected_2018,liu_locally_2000,wang_harnessing_2014,buckmann_mechanical_2015,misseroni_cymatics_2016,zhou_elastic_2008,zhang_asymmetric_2020,cummer_controlling_2016,guenneau_acoustic_2007,kaina_slow_2017,tallarico_edge_2017,willis_negative_2016,wen_effects_2011,mitchell_metaconcrete_2014,zhu_total-internal-reflection_2018,zhu_negative_2014,kaina_negative_2015,gonella_interplay_2009}.
These inhomogeneities at the micro-scale usually cause anisotropic responses, meaning that the mechanical response of a unit-cell depends on the direction of the applied load.
Furthermore, due to the constructive interaction of the inhomogeneities, for periodic meta-materials the anisotropy often persists at higher scales, so that large specimens made up of a huge number of unit cells still show different responses for different loading directions.
Although it may be unknown a priori if the same class of anisotropy is retained while going from the micro- to the macro- scale, it is a conservative hypothesis to assume that the macroscopic class of symmetry cannot be higher than the microscopic one.
In light of this, we choose the tetragonal symmetry class for all the elastic and micro-inertia tensors of the relaxed micromorphic model given the symmetries of the unit cell (see Fig.\ref{fig:fig_tab_unit_cell_0}).
Because of the anisotropy class chosen, the relaxed micromorphic parameters can be fitted just on two independent directions of propagation (for example 0 $^{\circ}$ and 45 $^{\circ}$), and the metamaterial's response will be retrieved for all the other intermediate directions without needing to change the value of the fitted parameters, which means that the parameters of the model are angle- independent.
Remarkably, the relaxed micromorphic parameters are also frequency-independent, which is another strong point of our homogenized model.
This is a major strength of the relaxed micromorphic model, since standard homogenization techniques do not allow to directly transfer anisotropy from the micro to the macro scale \cite{touboul2020effective,maurel2019multimodal}, and often give rise to frequency-dependent homogenization relations \cite{chen_dispersive_2001,willis_exact_2009,craster_high_frequency_2010,willis_effective_2011,willis_construction_2012,boutin_large_2014,sridhar_general_2018}.
\subsection{A periodic metamaterial for acoustic wave control}
\begin{figure}[H]
	\begin{subfigure}{0.59\textwidth}
		\centering
		\includegraphics[width=0.5\textwidth]{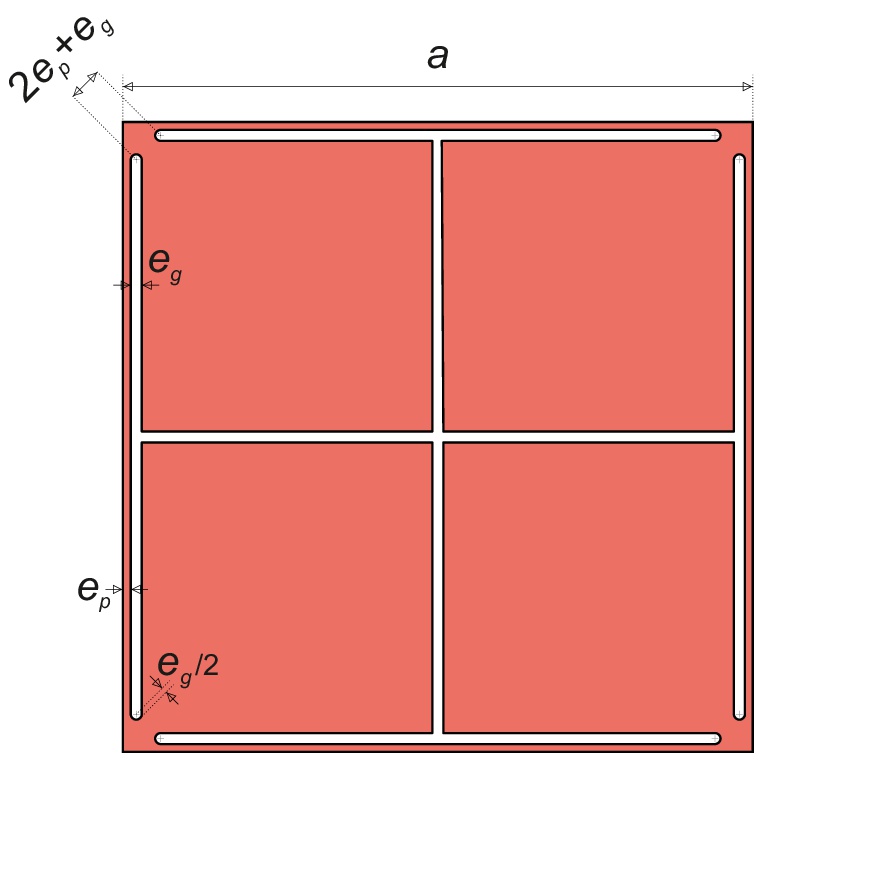}
	\end{subfigure}
	\hspace*{-1.5cm}
	\begin{minipage}{0.39\textwidth}
		\begin{subfigure}{\textwidth}
			\centering
			\vspace{0.6cm}
			\begin{tabular}{cccccc}
				\hline
				\textsl{a} & $e_{g}$ & $e_{p}$      \\ 
				$[$mm$]$ &  [mm]   &  [mm]   \\
				\hline
				20 & 0.35 & 0.25 \\ 
				\hline
				$\rho_{Ti}$ & $\lambda_{\tiny \mbox{Ti}}$ & $\mu_{\tiny \mbox{Ti}}$          \\
				$\mbox{[kg/m}^3\mbox{]}$ &   [GPa]  &    [GPa] \\
				\hline
				4400 & 88.8 & 41.8\\ 
				\hline
				\vspace{0.2cm}
			\end{tabular}
		\end{subfigure}
	\end{minipage}
	\caption{(\textit{left panel}) unit cell whose periodic repetition in space gives rise to the metamaterial treated in the present paper.
	(\textit{right panel}) Table of the geometry and material properties of the unit cell: $\rho_{\tiny \mbox{Ti}}$, $\lambda_{\tiny \mbox{Ti}}$, and $\mu_{\tiny \mbox{Ti}}$ stand for the density and the Lamé constants of titanium, respectively.}
	\label{fig:fig_tab_unit_cell_0}
\end{figure}

The unit cell (Fig.~\ref{fig:fig_tab_unit_cell_0}) used in this work has already been introduced and characterized in \cite{rizzi_exploring_2021,rizzi2020towards} and we recall its geometric and elastic properties in the table of Fig.~\ref{fig:fig_tab_unit_cell_0}.
This unit cell has been conceived for elastic waves control in the acoustic regime.
In fact, dispersion curves for the associated periodic metamaterials can be obtained via standard Bloch-Floquet analysis (black dotted line in Fig.~\ref{fig:disp_curve}) and show a band-gap between 1725 and 2136 Hz.

Standard Bloch-Floquet analysis is very effective to describe the metamaterials's behaviours when it is sufficiently large to be modelled as an infinite medium.
However, it is not suitable to describe the response of finite-size metamaterials.
Current approaches exploring the dynamic response of finite-size metamaterials are often limited to FEM simulations encoding all the details of the underlying microstructure (see e.g. \cite{krushynska2017coupling}).
While these simulations are very precise, they are also computationally expensive, so that the possibility of studying large samples is limited due to the non-linear increase of the computational cost with the size of the sample itself.

Dynamic homogenization techniques that are able to provide PDEs describing the macroscopic material's response  have been extensively studied \cite{willis_exact_2009,willis_effective_2011,willis_construction_2012,craster_high_frequency_2010,nolde_high_2011,sridhar_general_2018,pham_transient_2013,sridhar_homogenization_2016,sridhar_frequency_2020}.
Yet, fundamental difficulties arise when specimens of finite size are considered, especially in the context of establishing well-posed homogenized boundary conditions.
To overcome these difficulties, we introduce the relaxed micromorphic model together with its intrinsically well-posed boundary conditions.

\subsection{Relaxed micromorphic modeling of a tetragonal metamaterial}
In \cite{rizzi2021boundary}, we established a coherent boundary value problem for relaxed micromorphic continua enabling the correct description of elastic waves scattering at metamaterials' interfaces.
Thanks to the simplified structure of our homogenized model (free of unnecessary parameters), we are able to overcome computational difficulties while being able to reproduce fundamental properties such as band-gaps, dispersion and anisotropy.
This simplification enables us to explore complex metamaterials for elastic wave control in a way that would not be viable otherwise.
For example, in \cite{rizzi2021boundary} we designed a meta-structure capable of energy focusing which also acts as a shield, in \cite{rizzi_exploring_2021} we conceived acoustic screens and acoustic absorbers, and in \cite{rizzi2020towards} we designed mechanical diodes.

In the present paper we go further in the exploration of meta-structures for elastic wave control and we show how the simplified structure of the relaxed micromorphic model allows us to effectively design a shield device.
Moreover, we show that this shield can be investigated for increasing sizes without incurring in substantial increment of computational loads.
The study presented in this paper is a fundamental stepping stone in view of the creation of larger meta-structures that control elastic waves and eventually recover energy.

\begin{figure}[H]
	\centering
	\begin{minipage}[H]{0.49\textwidth}
		\includegraphics[width=\textwidth]{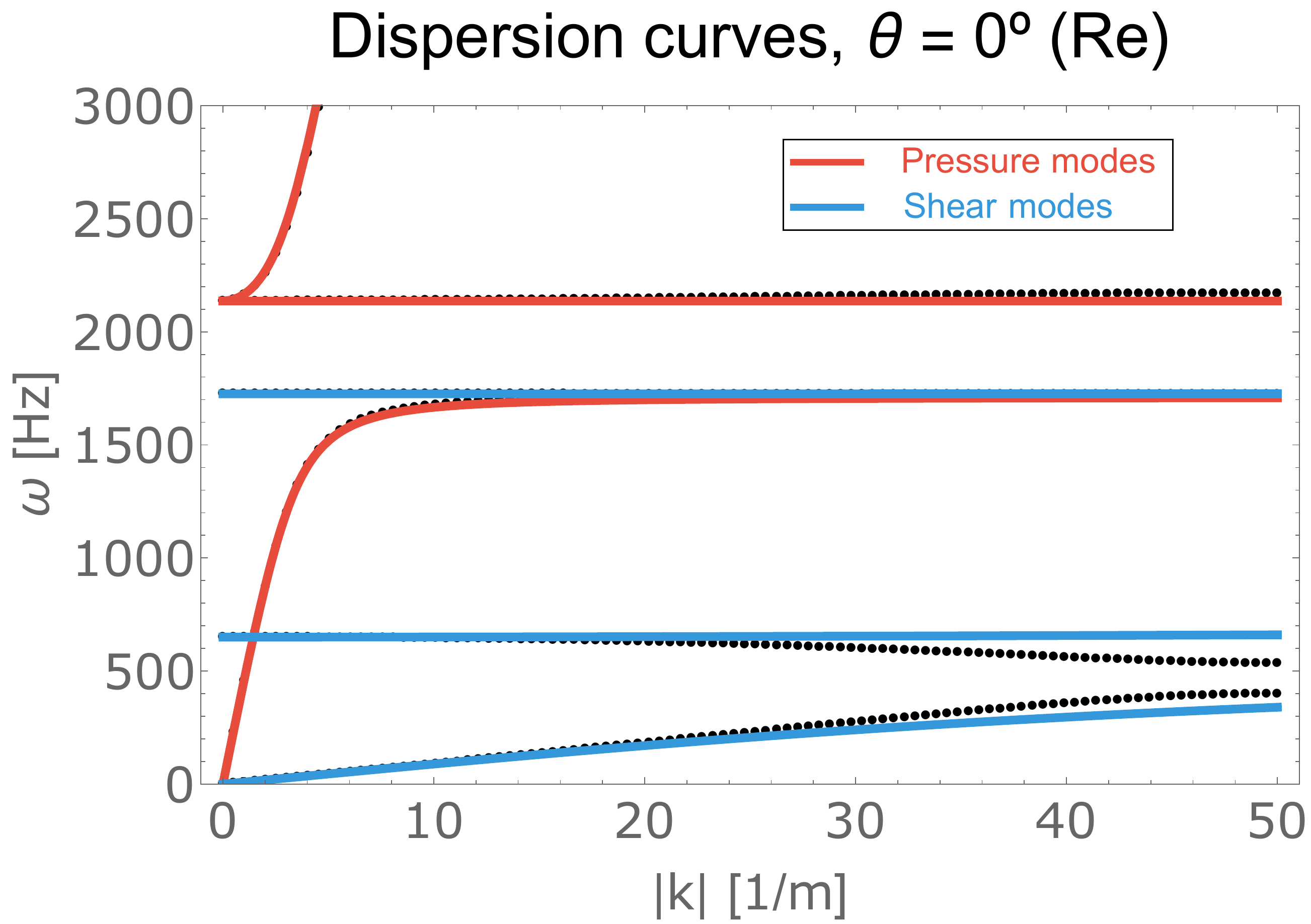}
	\end{minipage}
	\begin{minipage}[H]{0.49\textwidth}
		\includegraphics[width=\textwidth]{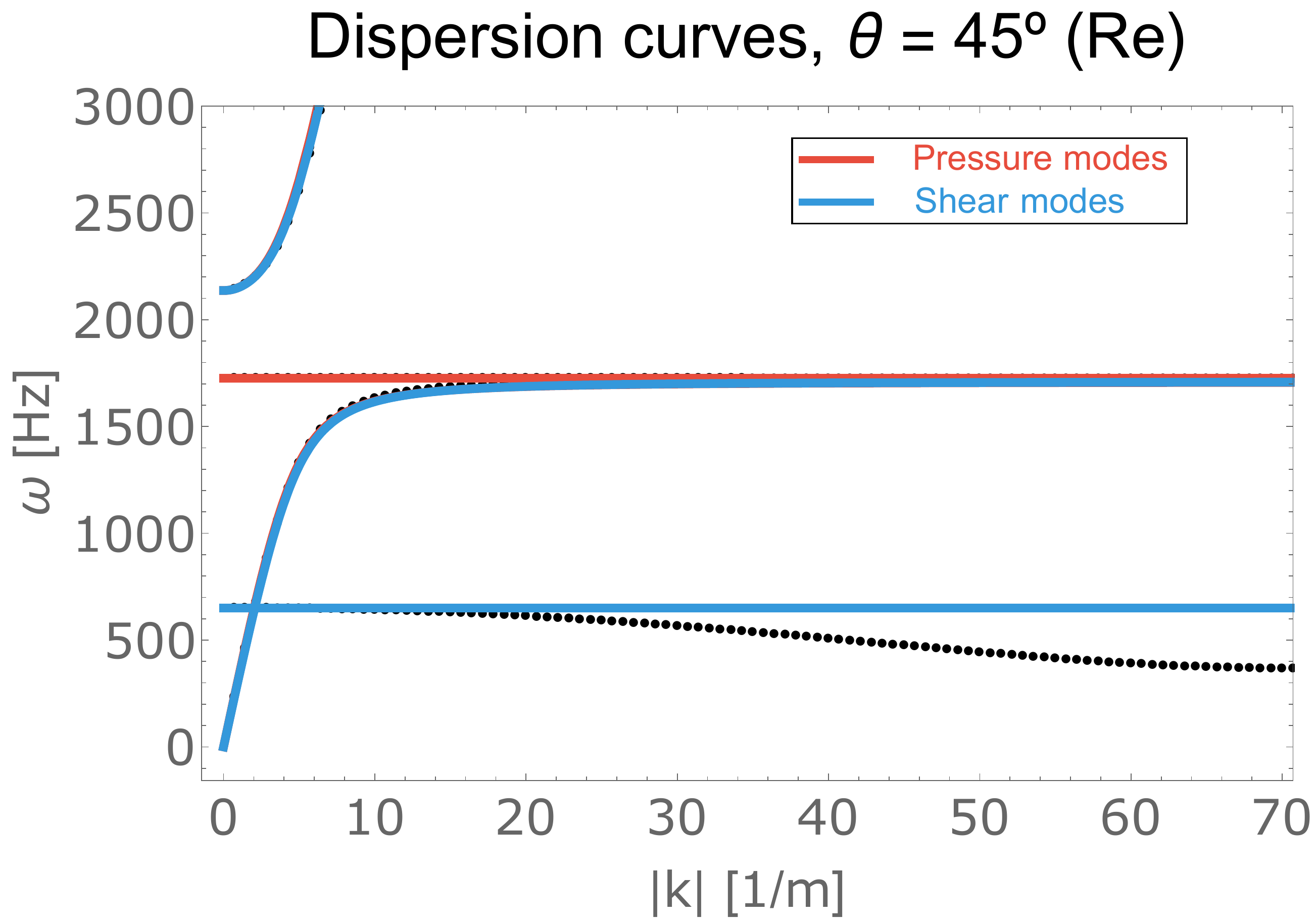}
	\end{minipage}
	\begin{minipage}[H]{0.49\textwidth}
		\includegraphics[width=\textwidth]{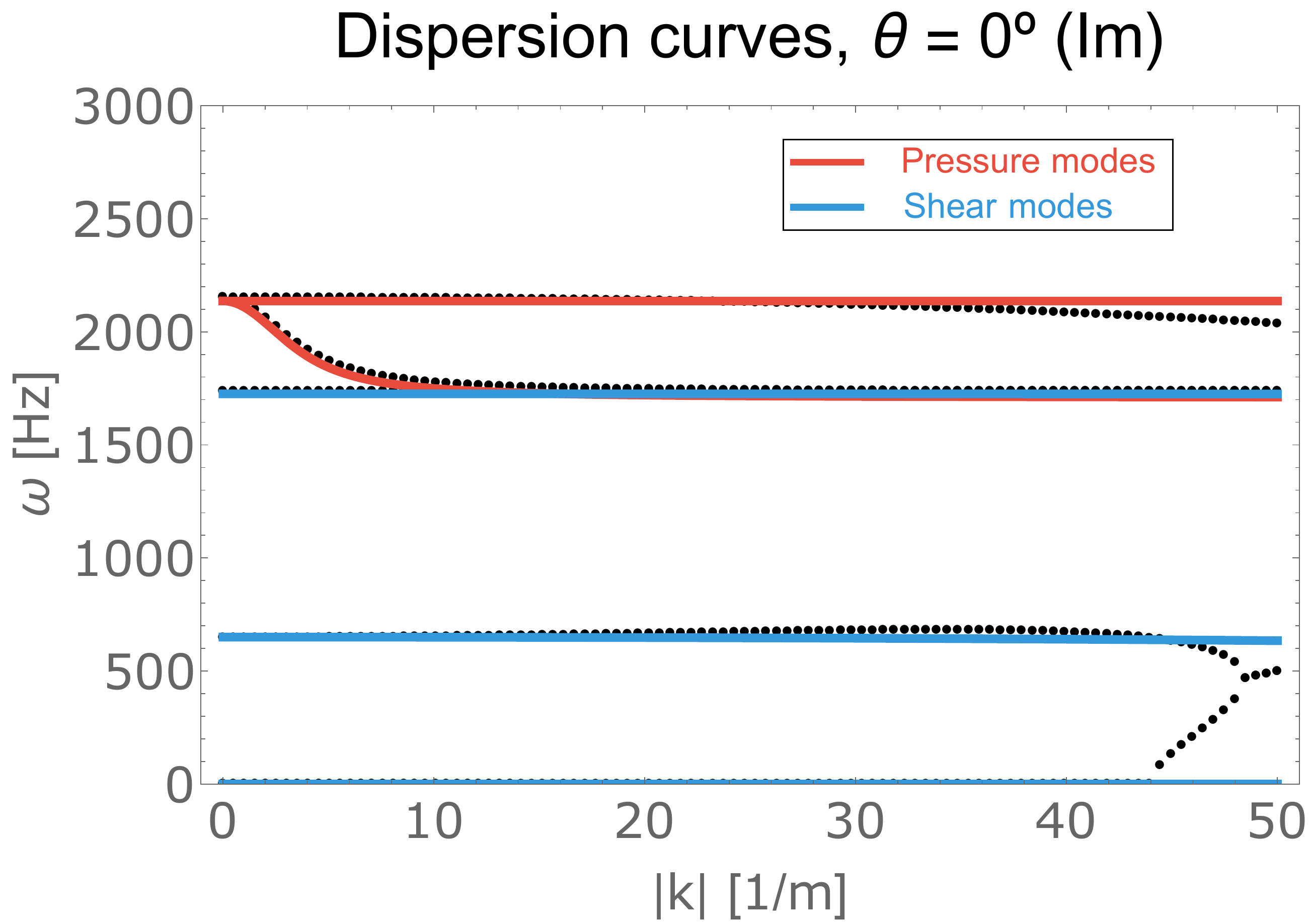}
	\end{minipage}
	\begin{minipage}[H]{0.49\textwidth}
		\includegraphics[width=\textwidth]{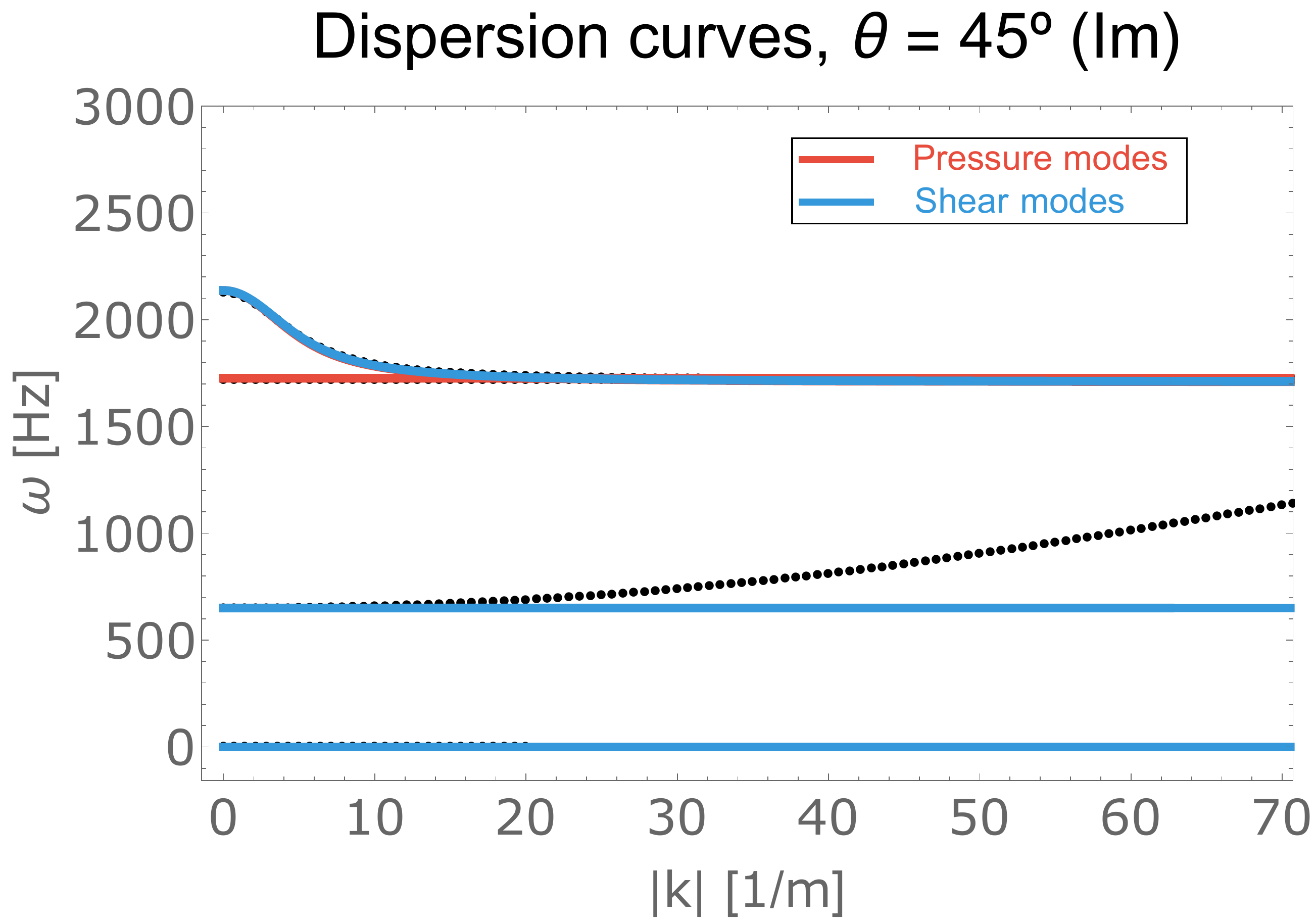}
	\end{minipage}
	\caption{
	Fitting of the real and imaginary components of the relaxed micromorphic dispersion curves (solid lines) on the Bloch-Floquet ones (dotted lines) for the unit cell shown in Fig~\ref{fig:fig_tab_unit_cell_0}.
	The fitting is carried out just for the real part of the dispersion curves and the imaginary part comes automatically.
	The fitting is performed using only two directions of propagation, namely 0$^{\circ}$ (\textit{left panels}) and 45$^{\circ}$ (\textit{right panels}).
	}
	\label{fig:disp_curve}
\end{figure}

We recall here that the kinetic and the strain energy densities for the relaxed micromorphic model (without curvature effects) are given by \cite{dagostino_effective_2020}
\!\!\!
\footnote{
The presence of curvature terms is essential to catch size-effects in the static regime that are not the target of the present paper.
Thus, we neglect higher order derivatives of the micro-distortion tensor in the expression of the strain energy density $W$.
}
\begin{equation}
\begin{array}{ll}
J \left(u_{,t},\nabla u_{,t}, P_{,t}\right) =& 
\dfrac{1}{2}\rho \, \langle u_{,t},u_{,t} \rangle + 
\dfrac{1}{2} \langle \mathbb{J}_{\text{micro}}  \, \mbox{sym} \, P_{,t}, \mbox{sym} \, P_{,t} \rangle 
+ \dfrac{1}{2} \langle \mathbb{J}_{\text{c}} \, \mbox{skew} \, P_{,t}, \mbox{skew} \, P_{,t} \rangle \\[3mm]
& + \dfrac{1}{2} \langle \mathbb{T}_{\text{e}} \, \mbox{sym}\nabla u_{,t}, \mbox{sym}\nabla u_{,t} \rangle
  + \dfrac{1}{2} \langle \mathbb{T}_{\text{c}} \, \mbox{skew}\nabla u_{,t}, \mbox{skew}\nabla u_{,t} \rangle,
\end{array}
\label{eq:kinEneMic}
\end{equation}

\begin{equation}
\begin{array}{ll}
W \left(\nabla u, P\right) =& 
  \dfrac{1}{2} \langle \mathbb{C}_{\text{e}} \, \mbox{sym}\left(\nabla u -  \, P \right), \mbox{sym}\left(\nabla u -  \, P \right) \rangle
+ \dfrac{1}{2} \langle \mathbb{C}_{\text{micro}} \, \mbox{sym}  \, P,\mbox{sym}  \, P \rangle\\[3mm]
& 
+ \dfrac{1}{2} \langle \mathbb{C}_{\text{c}} \, \mbox{skew}\left(\nabla u -  \, P \right), \mbox{skew}\left(\nabla u -  \, P \right) \rangle
\, ,
\end{array}
\label{eq:strainEneMic}
\end{equation}
where $P \in \mathbb{R}^{3\times3}$ is the non-symmetric micro-distortion tensor, $u$ is the macroscopic displacement field, $\rho$ is the macroscopic apparent density, and $\mathbb{J}_{\text{micro}}$, $\mathbb{J}_{\text{c}}$, $\mathbb{T}_{\text{e}}$, $\mathbb{T}_{\text{c}}$ are 4th order micro-inertia tensors, and $\mathbb{C}_{\text{e}}$, $\mathbb{C}_{\text{micro}}$, and $\mathbb{C}_{\text{c}}$ are 4th order elastic tensors.
This gives rise to the following equilibrium equations
\begin{equation}
\rho \, u_{,tt} - \mbox{Div}\left(\widehat{\sigma}_{,tt}\right) = \mbox{Div}\left(\widetilde{\sigma}\right) \, ,
\qquad
\left( \mathbb{J}_{\text{micro}} + \mathbb{J}_{\text{c}} \right) \, P_{,tt} = \widetilde{\sigma} - s \, ,
\label{eq:equiMic}
\end{equation}
where 
\begin{equation}
\begin{array}{cc}
\widehat{\sigma} \coloneqq \mathbb{T}_{\text{e}}~\mbox{sym} \nabla u + \mathbb{T}_{\text{c}}~\mbox{skew} \nabla u \, ,
\qquad
s \coloneqq \mathbb{C}_{\text{micro}}~\mbox{sym}  \, P \, ,
\\[3mm]
\widetilde{\sigma} \coloneqq \mathbb{C}_{\text{e}}~\mbox{sym}\left(\nabla u -  \, P \right) + \mathbb{C}_{\text{c}}~\mbox{skew}\left(\nabla u -  \, P \right) \, ,
\end{array}
\label{eq:sigMic}
\end{equation}
together with the associated boundary conditions \cite{rizzi2021boundary}
\!\!\!
\footnote{
Since we are neglecting higher order derivatives for the micro-distortion tensor, only a condition on the generalized traction is needed.
When considering curvature terms an extra condition on the so-called double-traction should be considered (see \cite{aivaliotis_frequency-_2020} for more details).
}
\begin{align}
t_{\text{m}} =
\left(\widetilde{\sigma} + \widehat{\sigma}_{,tt} \right) n
= f^{\text{ext}} \, ,
\label{eq:BC}
\end{align}

where $t_{\text{m}}$ is the generalized internal traction, $f^{\text{ext}}$ is the external traction, and $n$ is the normal to the boundary.
Given the tetragonal symmetry of the cell in Fig.~\ref{fig:fig_tab_unit_cell_0}, we consider that the metamaterial generated from this cell keeps the same (tetragonal) symmetry at the macroscopic scale.
In light of this, the elastic and micro-inertia tensors appearing in eq.(\ref{eq:kinEneMic}) and eq.(\ref{eq:strainEneMic}) take the tetragonal form

\begin{equation}
\begin{array}{rlrl}
	\mathbb{J}_{\text{micro}} &=
	\begin{pmatrix}
	\eta_{3} + 2\eta_{1} & \eta_{3}              & \dots 			& \bullet \\ 
	\eta_{3}            & \eta_{3} + 2\eta_{1} & \dots 			& \bullet \\ 
	\vdots               & \vdots                 & \ddots 			& \bullet \\ 
	\bullet              & \bullet                & \bullet 		& \eta^{*}_{1} \\ 
	\end{pmatrix} \, ,
	\quad
	&\mathbb{J}_{\text{c}} &=
	\begin{pmatrix}
	\bullet & 			& \bullet\\ 
	& \ddots 	& \vdots\\ 
	\bullet & \dots & 4\eta_{2}
	\end{pmatrix} \, ,
	\\[1.5cm]
	\mathbb{T}_{\text{e}} &=
	\begin{pmatrix}
	\overline{\eta}_{3} + 2\overline{\eta}_{1}	& \overline{\eta}_{3}        			   	& \dots		& \bullet\\ 
	\overline{\eta}_{3}         				    & \overline{\eta}_{3} + 2\overline{\eta}_{1} 	& \dots		& \bullet\\ 
	\vdots                    					& \vdots                 				   	& \ddots 	&		 \\ 
	\bullet                   					& \bullet									& 	 		& \overline{\eta}^{*}_{1}
	\end{pmatrix} \, ,
	\quad
	&\mathbb{T}_{\text{c}} &=
	\begin{pmatrix}
	\bullet & 			& \bullet\\ 
	& \ddots 	& \vdots\\ 
	\bullet & \dots & 4\overline{\eta}_{2}
	\end{pmatrix} \, ,
    \\[1.5cm]
    \mathbb{C}_{\text{e}} &= 
    \begin{pmatrix}
    \lambda_{\text{e}} + 2\mu_{\text{e}}	& \lambda_{\text{e}}				& \dots		& \bullet\\ 
    \lambda_{\text{e}}				& \lambda_{\text{e}} + 2\mu_{\text{e}}	& \dots		& \bullet\\ 
    \vdots					& \vdots					& \ddots	& 		 \\ 
    \bullet					& \bullet					& 			& \mu_{\text{e}}^{*}\\ 
    \end{pmatrix} \, ,
    \quad
    &\mathbb{C}_{\text{c}} &= 
    \begin{pmatrix}
    \bullet & 			& \bullet\\ 
    		& \ddots 	& \vdots\\ 
    \bullet & \dots		& 4\mu_{\text{c}}
    \end{pmatrix} \, ,
    \\[1.5cm]
    \mathbb{C}_{\text{micro}} &= 
    \begin{pmatrix}
    \lambda_{\text{micro}} + 2\mu_{\text{micro}}	& \lambda_{\text{micro}}				& \dots		& \bullet\\ 
    \lambda_{\text{micro}}				& \lambda_{\text{micro}} + 2\mu_{\text{micro}}	& \dots		& \bullet\\ 
    \vdots					& \vdots					& \ddots	&  \\ 
    \bullet					& \bullet					& 			& \mu_{\text{micro}}^{*}\\ 
    \end{pmatrix} \, .
    \end{array}
\label{eq:micro_ine_1}
\end{equation}
Following the fitting procedure given in \cite{rizzi_exploring_2021}, the relaxed micromorphic parameters are fitted on the metamaterial in Fig.~\ref{fig:fig_tab_unit_cell_0} and take the values given in Table~\ref{tab:parameters_RM_0}.
\begin{table}[H]
    \renewcommand{\arraystretch}{1.2}
    \centering
    \begin{subtable}[t]{.55\textwidth}
    \centering
    \begin{tabular}{cccccc}
        \\
        \thickhline
        \thickhline
        & $\rho$ & $\mu_{\text{e}}$ & $\lambda_{\text{e}}$ & $\mu_{\text{e}}^\star$ &
        \\
        & [\text{kg/m}$^3$] & [Pa] & [Pa] & [Pa] &
        \\
        \hline
        & 3841 & 2.53$\cross 10^9$ & 1.01$\cross 10^8$ & 1.26 $\cross 10^{6}$ &
        \\
        \thickhline
        & $\mu_{\text{micro}}$ & $\lambda_{\text{micro}}$ & $\mu_{\text{micro}}^\star$ & $\mu{\text{c}}$ &
        \\
        & [Pa] & [Pa] & [Pa] & [Pa] &
        \\
        \hline
        & 4.51 $\cross 10^{9}$ & 1.83 $\cross 10^{8}$ & 2.70 $\cross 10^{8}$ & $10^{5}$ &
        \\
        \thickhline
        & $\eta_1$ & $\eta_2$ & $\eta_3$ & $\eta_1^\star$ &
        \\
        & [\text{kg/m}] & [\text{kg/m}] & [\text{kg/m}] & [\text{kg/m}] &
        \\
        \hline
        & 38.99 & 5.99$\cross 10^{-3}$ & 1.58 & 2.31 &
        \\
        \thickhline
        & $\overline{\eta}_1$ & $\overline{\eta}_2$ & $\overline{\eta}_3$ & $\overline{\eta}^\star_1$ &
        \\
        & [\text{kg/m}] & [\text{kg/m}] & [\text{kg/m}] & [\text{kg/m}] &
        \\
        \hline
        & 8$\cross 10^{-4}$ & 0.02 & 0.008 & 0.09 &
        \\
        \thickhline
        \thickhline
    \end{tabular}
    \end{subtable}
    \hfill
	\centering
	\begin{subtable}[t]{.44\textwidth}
    \centering
    \begin{tabular}{ccccc}
        \thickhline
        \thickhline
    	& $\lambda_{\tiny \text{macro}}$ & $\mu_{\tiny \text{macro}}$ & $\mu^{*}_{\tiny \text{macro}}$ &
    	\\
    	& [Pa] & [Pa] & [Pa] &
    	\\
    	\hline
    	& $6.51\times10^7$ & $1.62\times10^9$ & $1.25\times10^6$ &
    	\\
    	\thickhline
    	\thickhline
    \end{tabular}
	\end{subtable}
    \caption{
    (\textit{left panel})
    Values of the elastic and micro-inertia parameters for the relaxed micromorphic material issued from the unit cell reported in Fig.~\ref{fig:fig_tab_unit_cell_0}, and 
    (\textit{right panel})
    the corresponding long-wave limit Cauchy material.
    }
     \label{tab:parameters_RM_0}
\end{table}
Assuming the harmonic wave ansatz, it is possible to derive from the equilibrium equation (\ref{eq:equiMic}) the relaxed micromorphic dispersion relations which are then fitted on the dispersion curves obtained for the microstructure with a Bloch-Floquet analysis \cite{aivaliotis_frequency-_2020,dagostino_effective_2020,rizzi_exploring_2021}.
The fitting of the dispersion curves obtained via the relaxed micromorphic model on those issued via Bloch-Floquet analysis is given in Fig.\ref{fig:disp_curve} both for the real and the imaginary part.
It is very important to note that the fitting procedure presented in \cite{dagostino_effective_2020,neff_identification_2020,rizzi_exploring_2021} is done only on the real part of the dispersion curves.
However, due to the consistency of the relaxed micromorphic model, the fitting of the imaginary part is retrieved automatically (it is not necessary to adjust the parameters or introduce additional ones).
On the same line, we underline that the aforementioned fitting procedure only relies on the fitting of the dispersion curves at 0$^{\circ}$ and 45$^{\circ}$, since, given the tetragonal symmetry class of the constitutive tensors used, these two independent directions are sufficient to determine all the parameters and automatically cover all the other directions.
These properties (self-given fitting of the imaginary part and of the intermediate directions) give further standing to our modelling approach, since, to the authors' knowledge, no model correctly describing metamaterials' behaviours for all possible directions and for a wide range of frequencies can be found in the literature.

Finally, we remark that Fig.~\ref{fig:disp_curve} also explicitly indicates a pressure/shear decomposition of the propagating relaxed micromorphic waves.
This decomposition is obtained by writing eqs.(\ref{eq:equiMic}) in a reference system aligned with the direction of propagation considered (0$^{\circ}$ and 45$^{\circ}$ respectively).
In fact, given the correct choice of the reference system, eqs.(\ref{eq:equiMic}) give rise to two sets of uncoupled PDEs that can be interpreted as pressure and shear waves.
Given the chosen class of symmetry (tetragonal), this uncoupling is only possible for the 0$^{\circ}$ and 45$^{\circ}$ direction, while waves are coupled when considering intermediate directions.
\section{Design of an acoustic shield}
\label{sec:comparison}
\begin{figure}[H]
	\centering
    \includegraphics[width=0.32\textwidth]{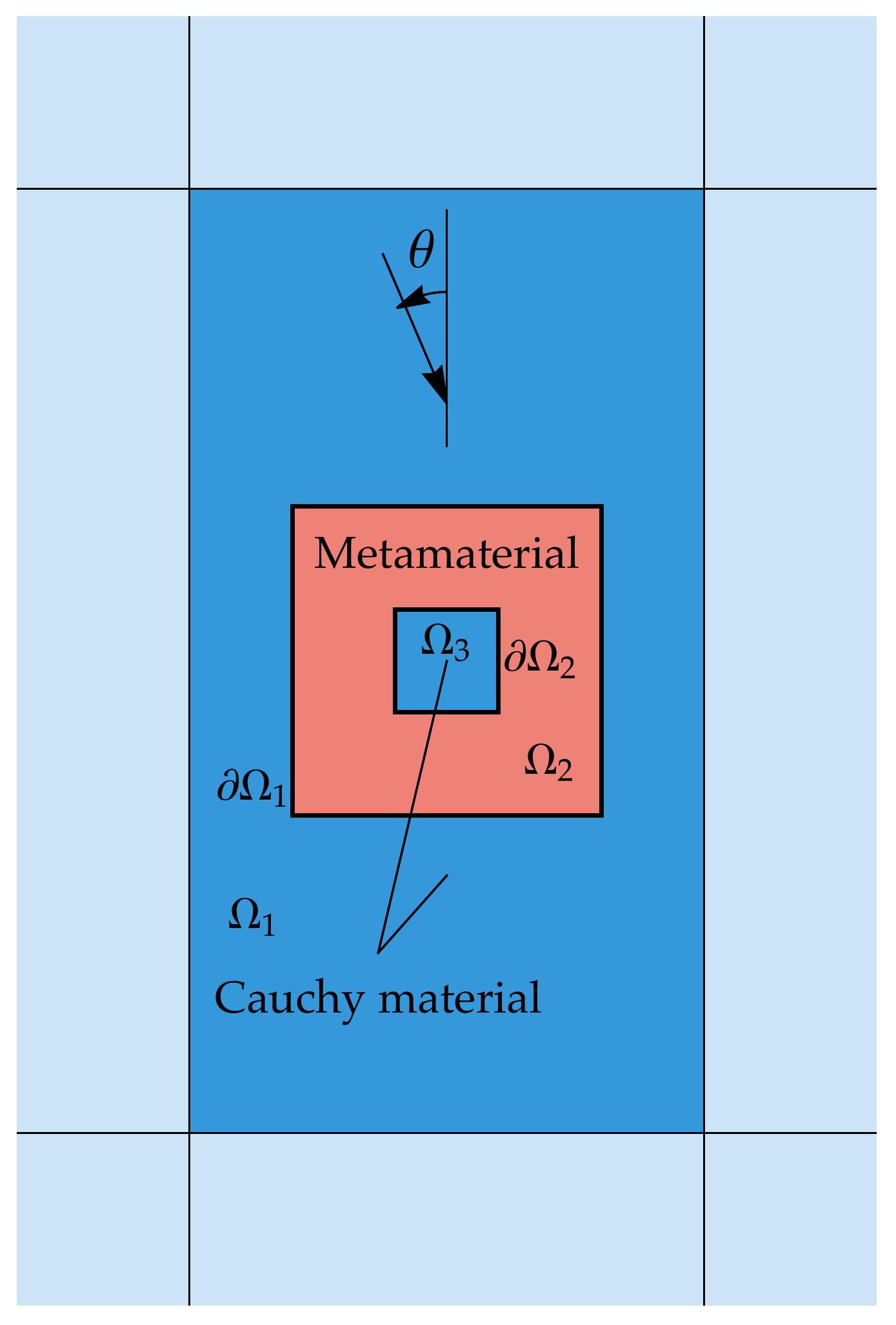}
	\caption{
	Schematic representation of the shield structure: we build a metamaterial frame ($\Omega_2$ in red) around an inclusion ($\Omega_3$ in blue) made up of the same Cauchy material found outside it ($\Omega_1$ in blue), in order to shield the inclusion from the external solicitation.
	The light blue domain represents the infinitely extended outer material.
	}
	\label{fig:Stru_Sche}
\end{figure}

We present here a comparison between the finite element modeling of the response of a structure (see Fig.~\ref{fig:Stru_Sche}) built out from the metamaterial composed by the unit cell in Fig.~\ref{fig:fig_tab_unit_cell_0} and the same structure made up of its homogeneous equivalent relaxed micromorphic continuum whose parameters are reported in Table~\ref{tab:parameters_RM_0}.
We consider different values of the frequency and of the direction of propagation of the incident wave.
\subsection{Implementing relaxed micromorphic and direct FEM simulations}
The results of the finite element analyses are obtained for the \textit{time harmonic} ansatz (plane incident wave) together with the following interface continuity conditions on the displacement and on the traction at $\partial \Omega_{i}$ (with $i=1,2$):
\begin{align}
\left\{
\begin{array}{rrrrrrrrrr}
u^{\text{c}} 
& = 
& u^{\text{r}} \, ,
\\*[2mm]
 t^{\text{c}} 
& = 
& t^{\text{r}} \, ,
\end{array}
\right.
\quad \mbox{on} 
\quad \partial \Omega_{i} \, ,
\label{eq:trac_1}
\end{align}
where $u^{\text{c}}$, $t^{\text{c}}$, $u^{\text{r}}$, and $t^{\text{r}}$ are the displacements and the tractions on the interfaces $\partial \Omega_{i}$.

For the simulations in which the frame is made up of the effective relaxed micromorphic material, the tractions $t^{\text{c}}$ and $t^{\text{r}}$ are computed as the limit from the Cauchy and the relaxed micromorphic side, respectively, and we recall again their expressions
\begin{align}
t^{\text{c}} = \sigma \, n \, ,
\qquad\qquad\qquad
t^{\text{r}} = \left(\widetilde{\sigma} + \widehat{\sigma}_{,tt} \right) n \, .
\end{align}

For the micro-structured material simulations, the conditions in eq.(\ref{eq:trac_1}) are identically satisfied on the boundaries $\partial \Omega_1$ and $\partial \Omega_2$, since the outside and inside material is the same that made up the matrix of the unite cell, so the continuity is automatically verified.
The only condition that must be verified is the traction-free conditions along the boundary of the cavity inside the unit cell.
\subsection{Results: broadband anisotropic relaxed micromorphic modeling of the acoustic shield}
\label{sec:comparison2}
\begin{figure}[H]
	\centering
	\begin{minipage}[H]{0.32\textwidth}
		\includegraphics[width=\textwidth]{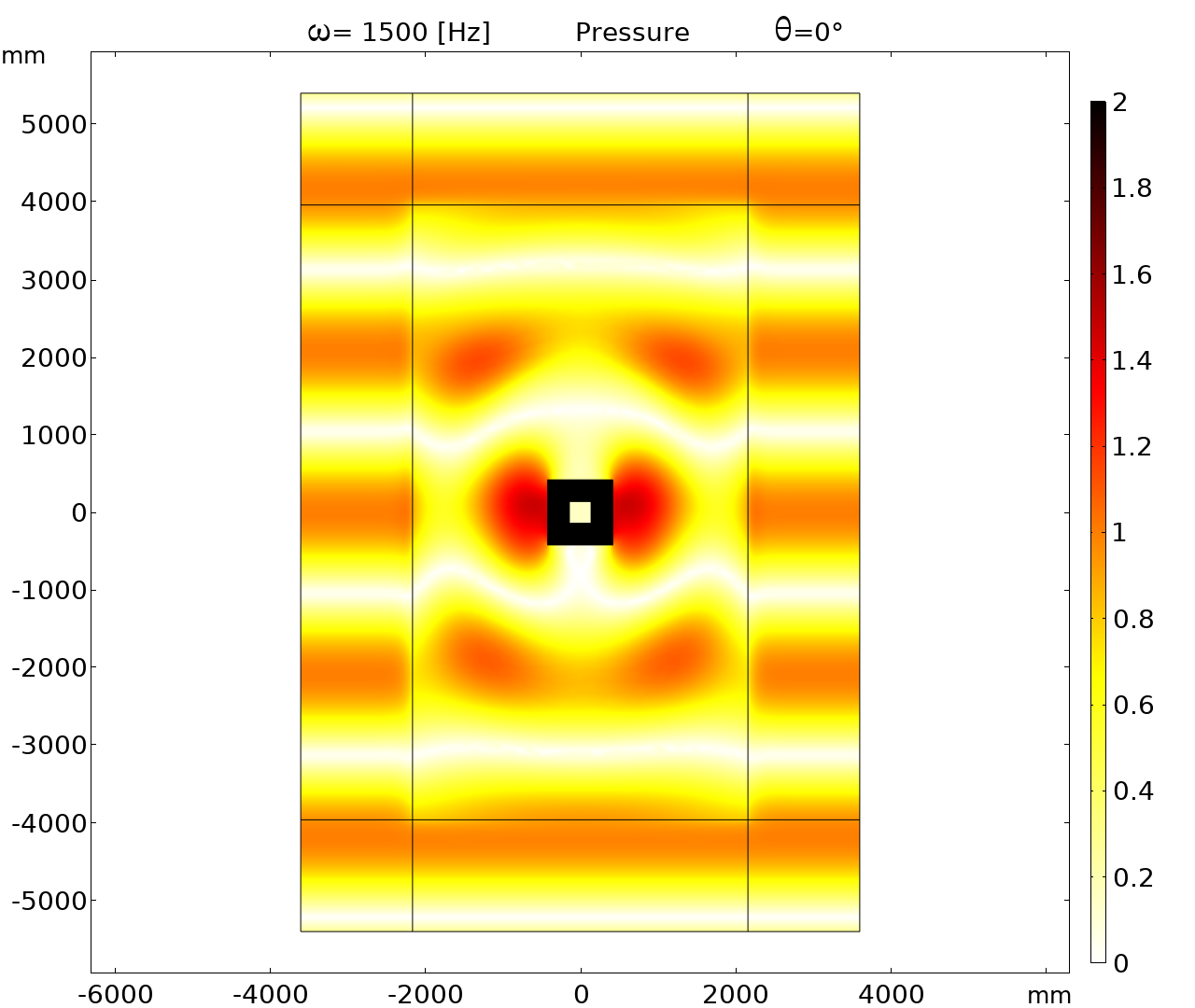}
	\end{minipage}
	\begin{minipage}[H]{0.32\textwidth}
		\includegraphics[width=\textwidth]{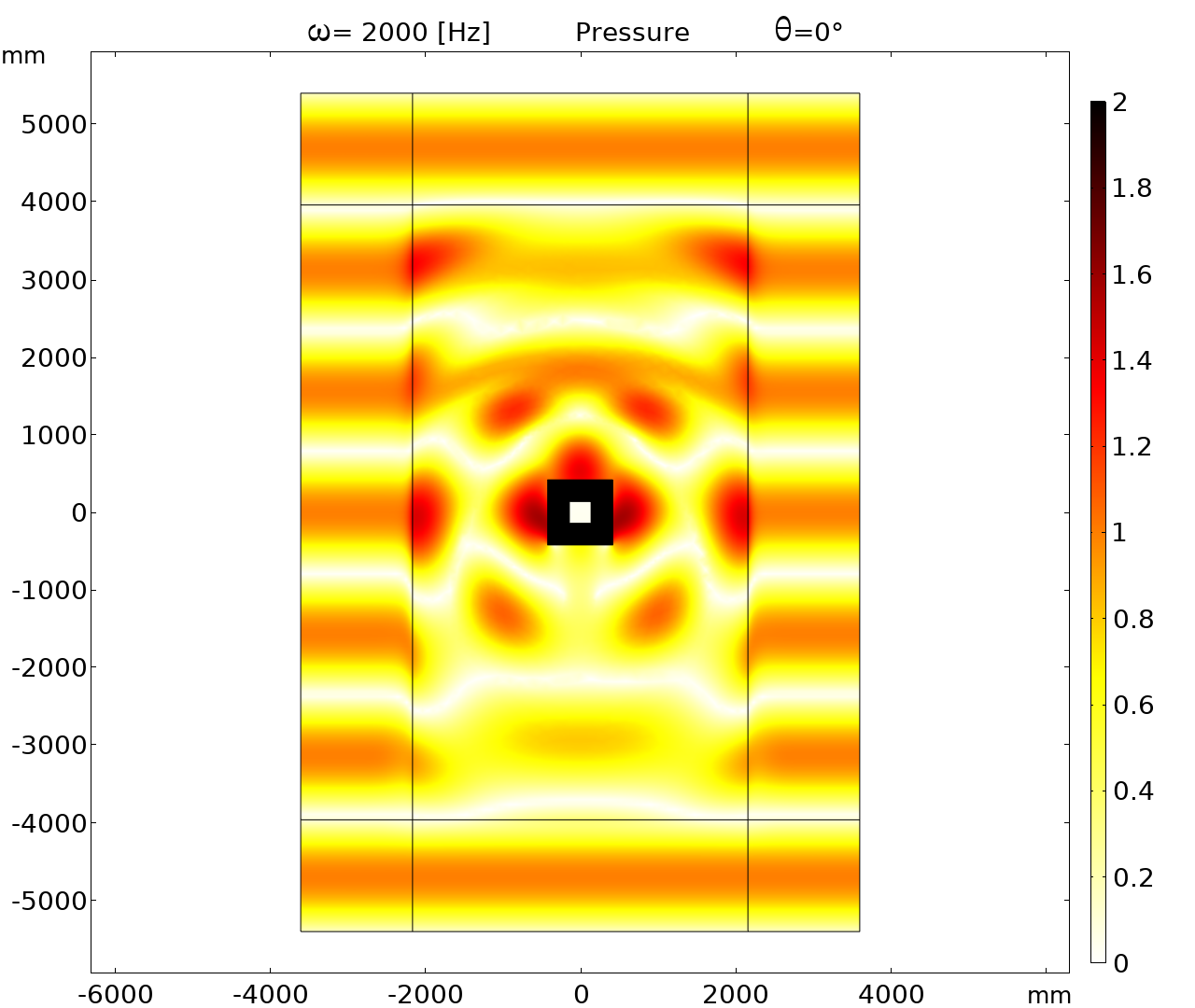}
	\end{minipage}
	\begin{minipage}[H]{0.32\textwidth}
		\includegraphics[width=\textwidth]{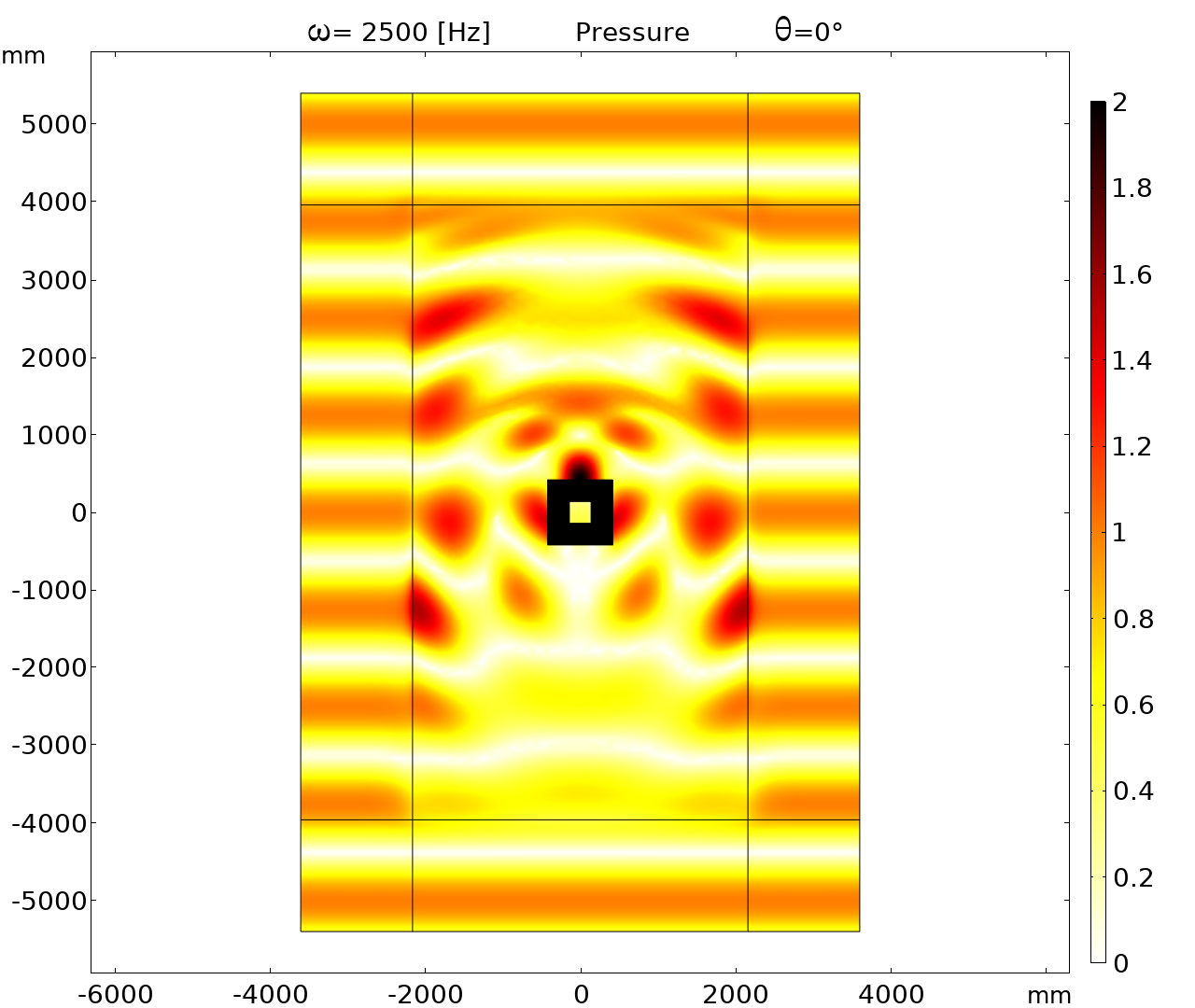}
	\end{minipage}
	\begin{minipage}[H]{0.32\textwidth}
		\includegraphics[width=\textwidth]{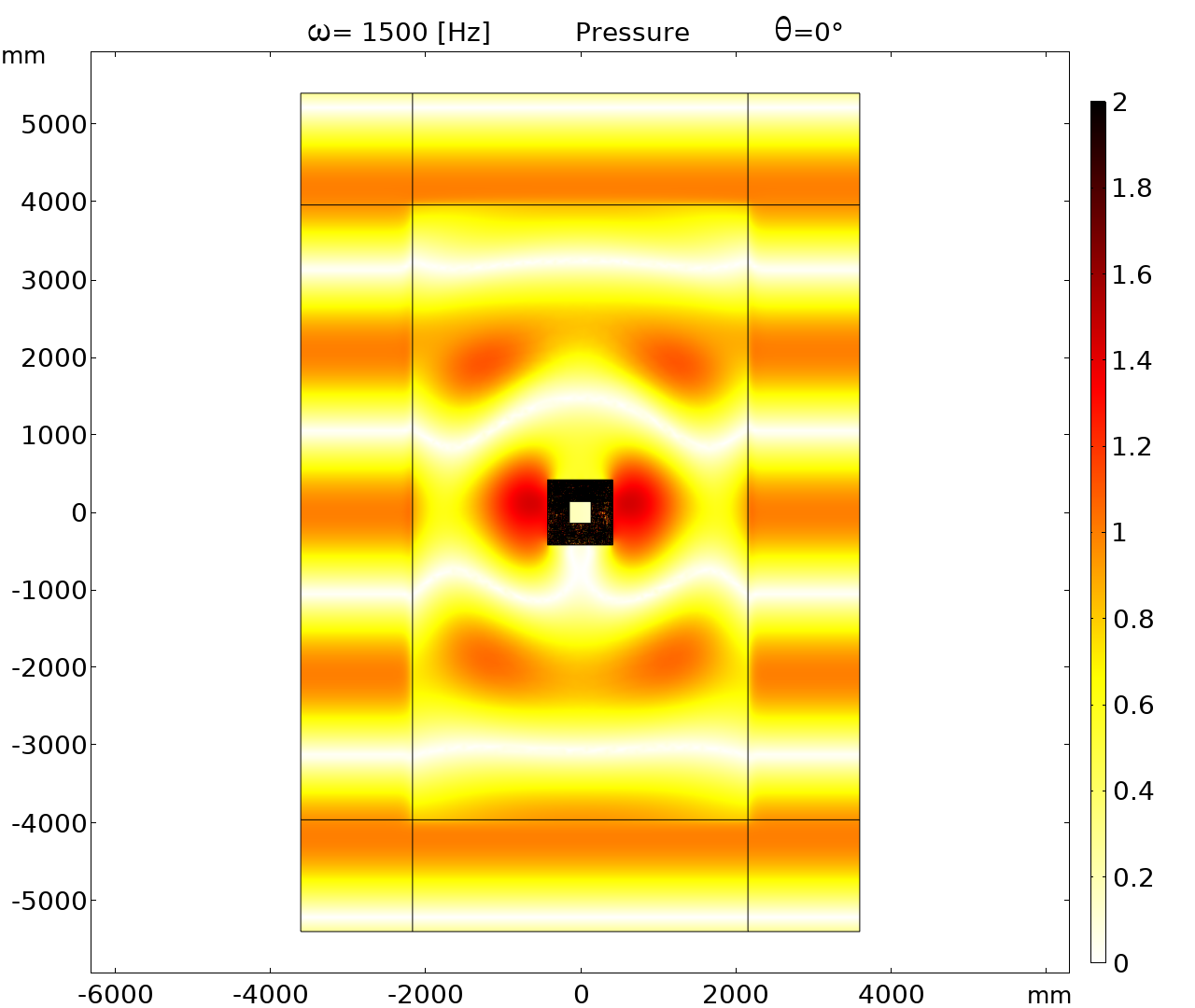}
	\end{minipage}
	\begin{minipage}[H]{0.32\textwidth}
		\includegraphics[width=\textwidth]{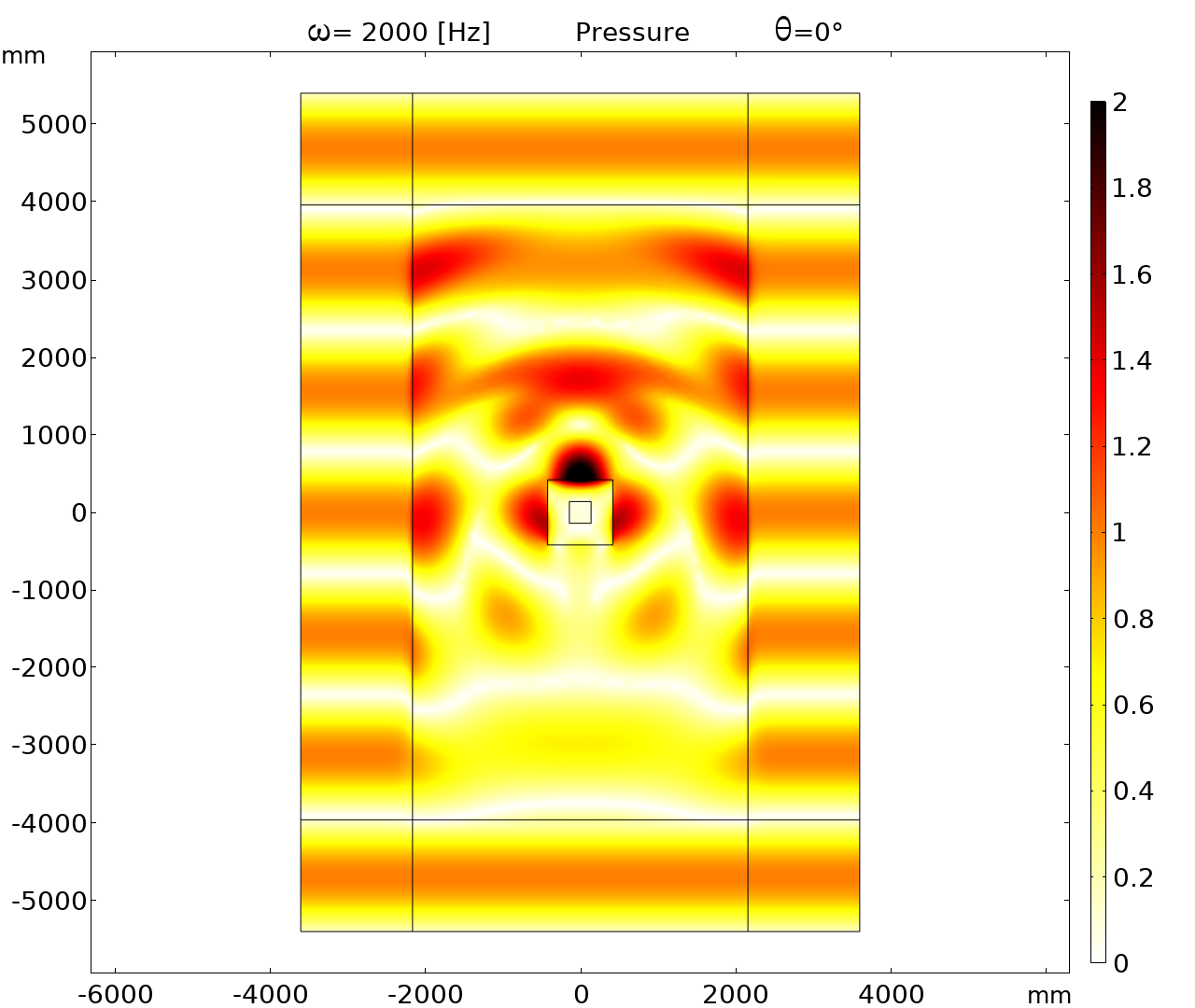}
	\end{minipage}
	\begin{minipage}[H]{0.32\textwidth}
		\includegraphics[width=\textwidth]{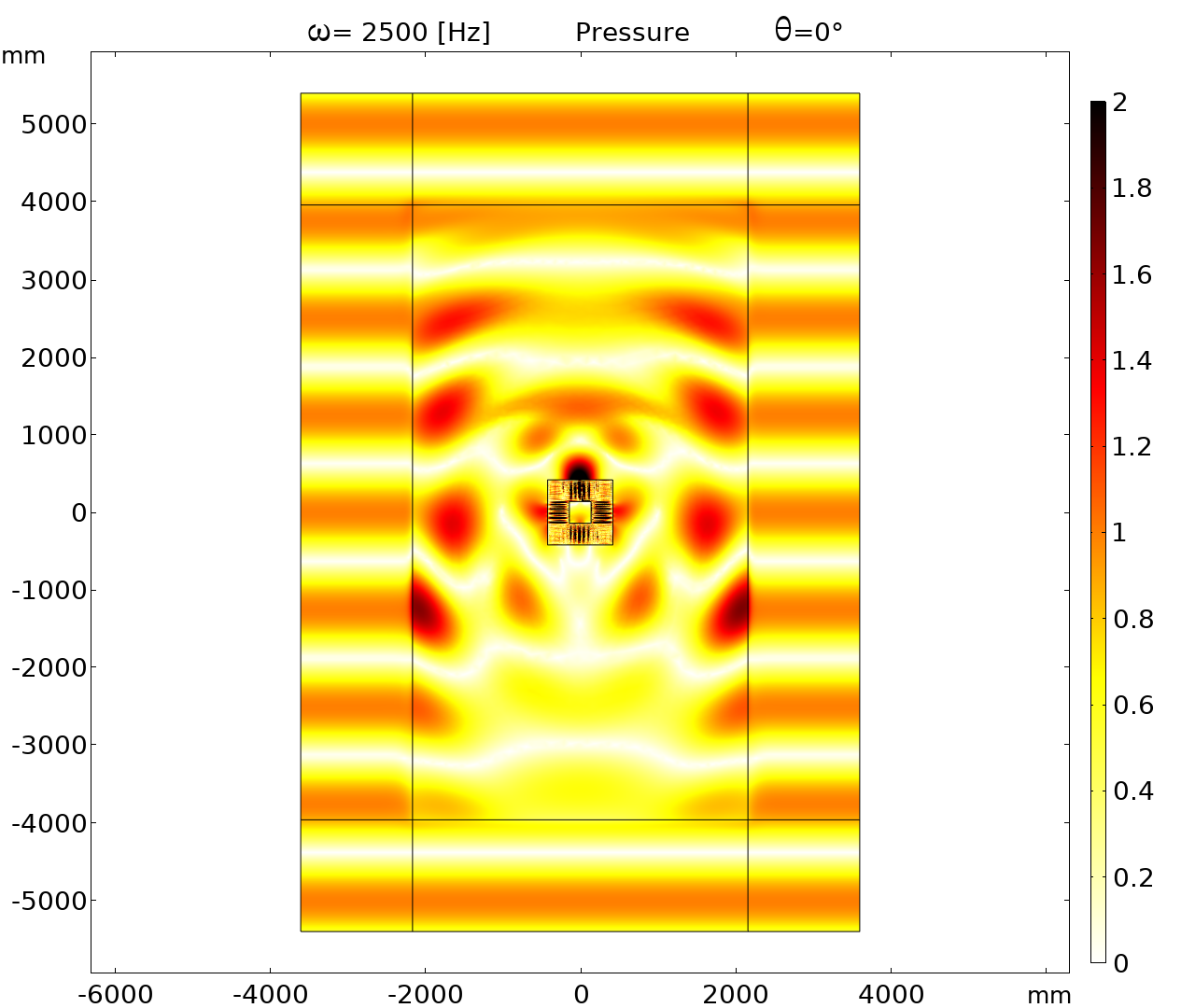}
	\end{minipage}
	\caption{
	Pressure incident wave's scattered field for a square made up of classic Cauchy material of side of 14 unit cells surrounded with a 14 unit cells thick metamaterial frame acting as a shield in the band-gap range.
	The angle of incidence is 0$^{\circ}$ and three frequencies are used.
	(\textit{top row}) Microstructured simulation.
	(\textit{bottom row}) Relaxed micromorphic simulation.
	(\textit{from left to right}) Frequency of 1.5 kHz (below the band-gap), 2 kHz (in the band-gap), and 2.5 kHz (above the band-gap).
	}
	\label{fig:Coll_compa_p_00}
\end{figure}

\begin{figure}[H]
	\centering
	\begin{minipage}[H]{0.32\textwidth}
		\includegraphics[width=\textwidth]{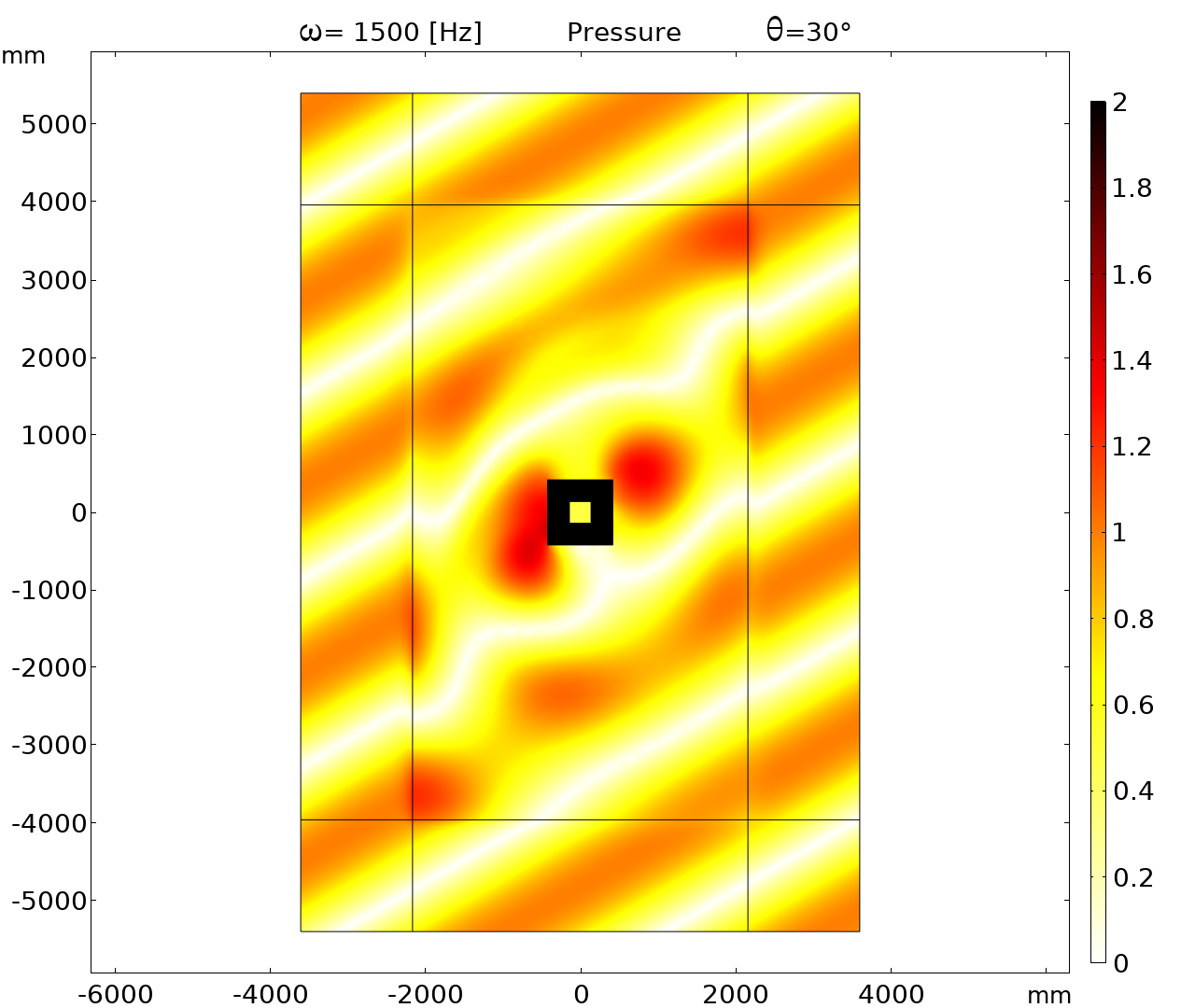}
	\end{minipage}
	\begin{minipage}[H]{0.32\textwidth}
		\includegraphics[width=\textwidth]{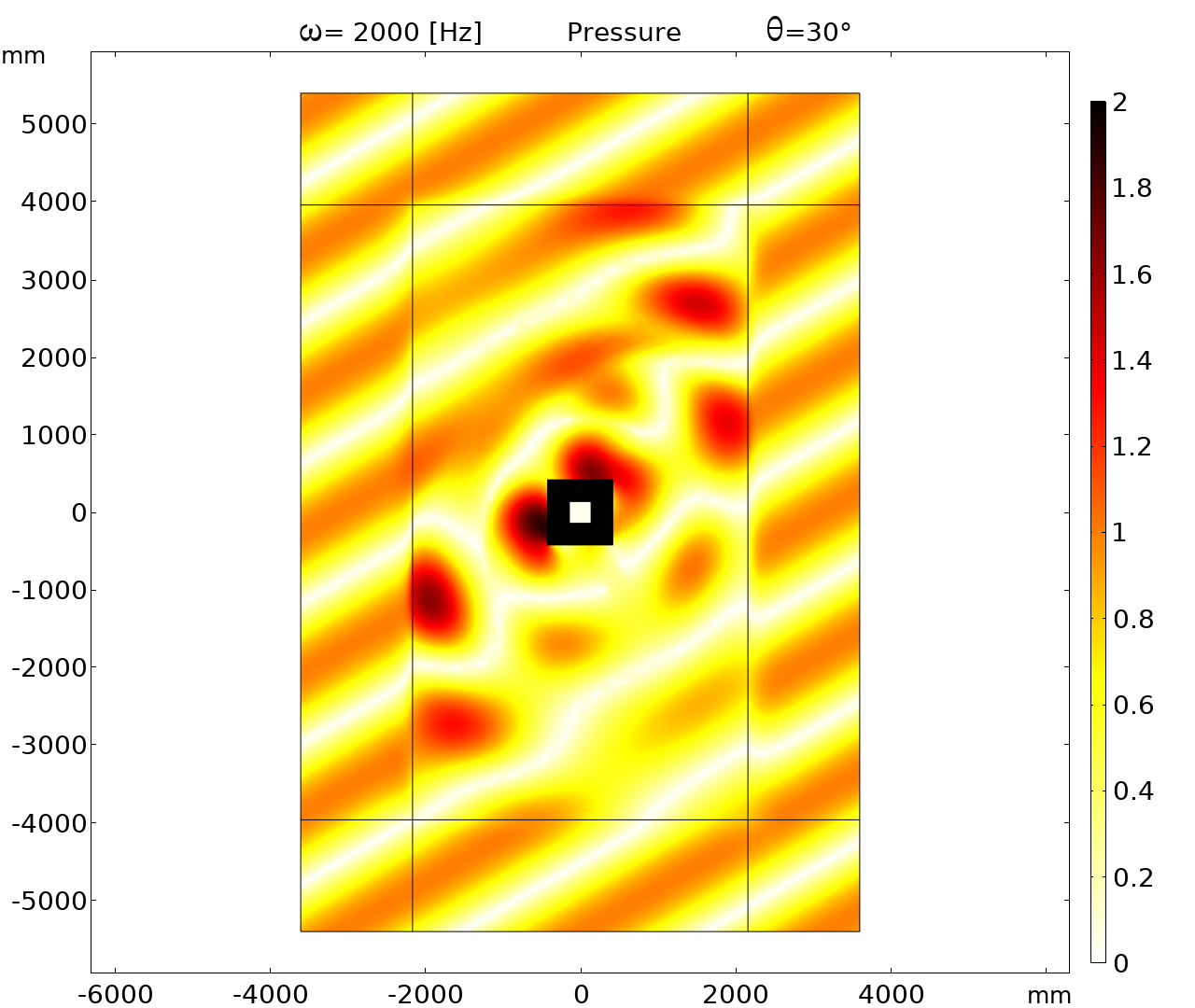}
	\end{minipage}
	\begin{minipage}[H]{0.32\textwidth}
		\includegraphics[width=\textwidth]{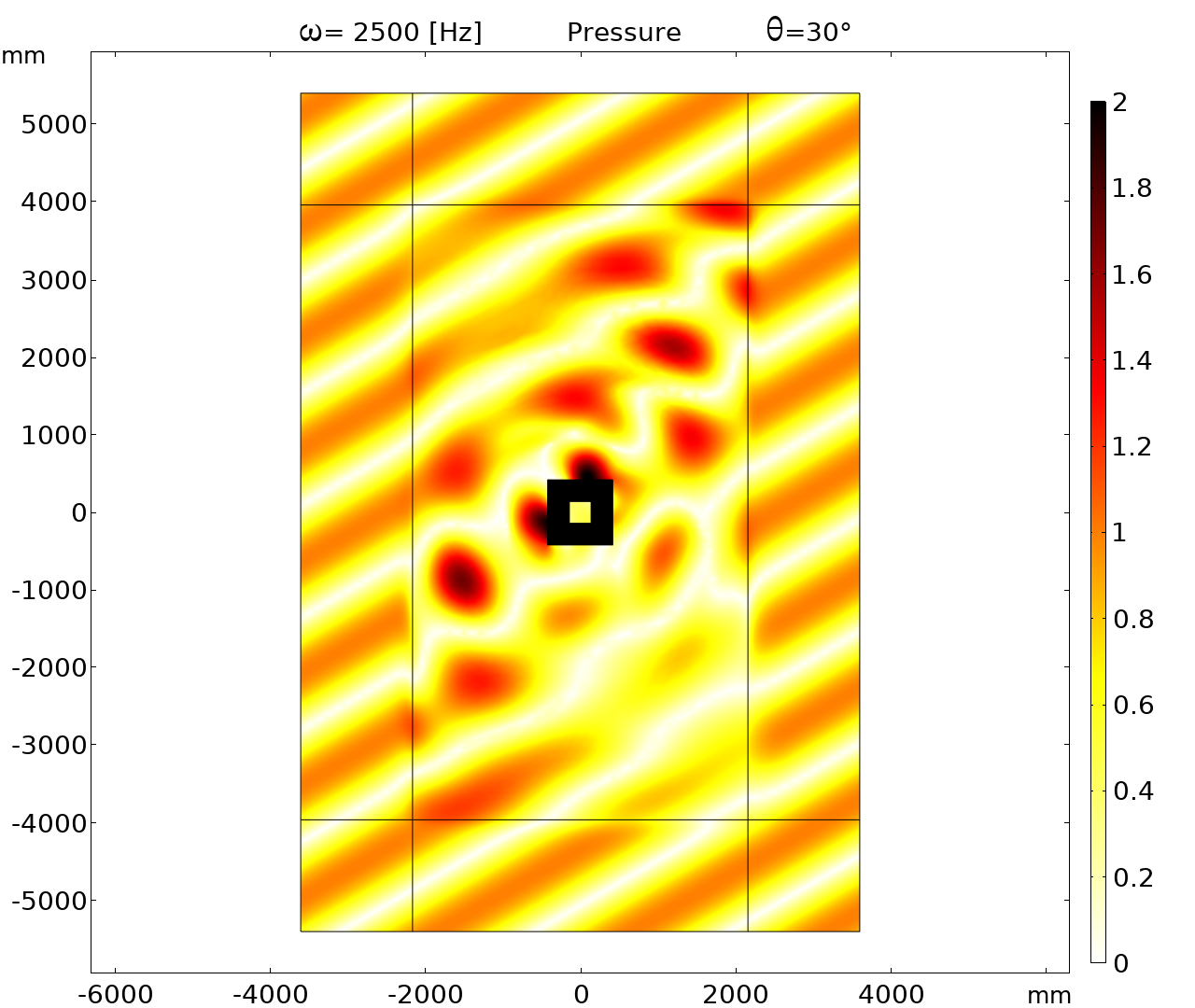}
	\end{minipage}
	\begin{minipage}[H]{0.32\textwidth}
		\includegraphics[width=\textwidth]{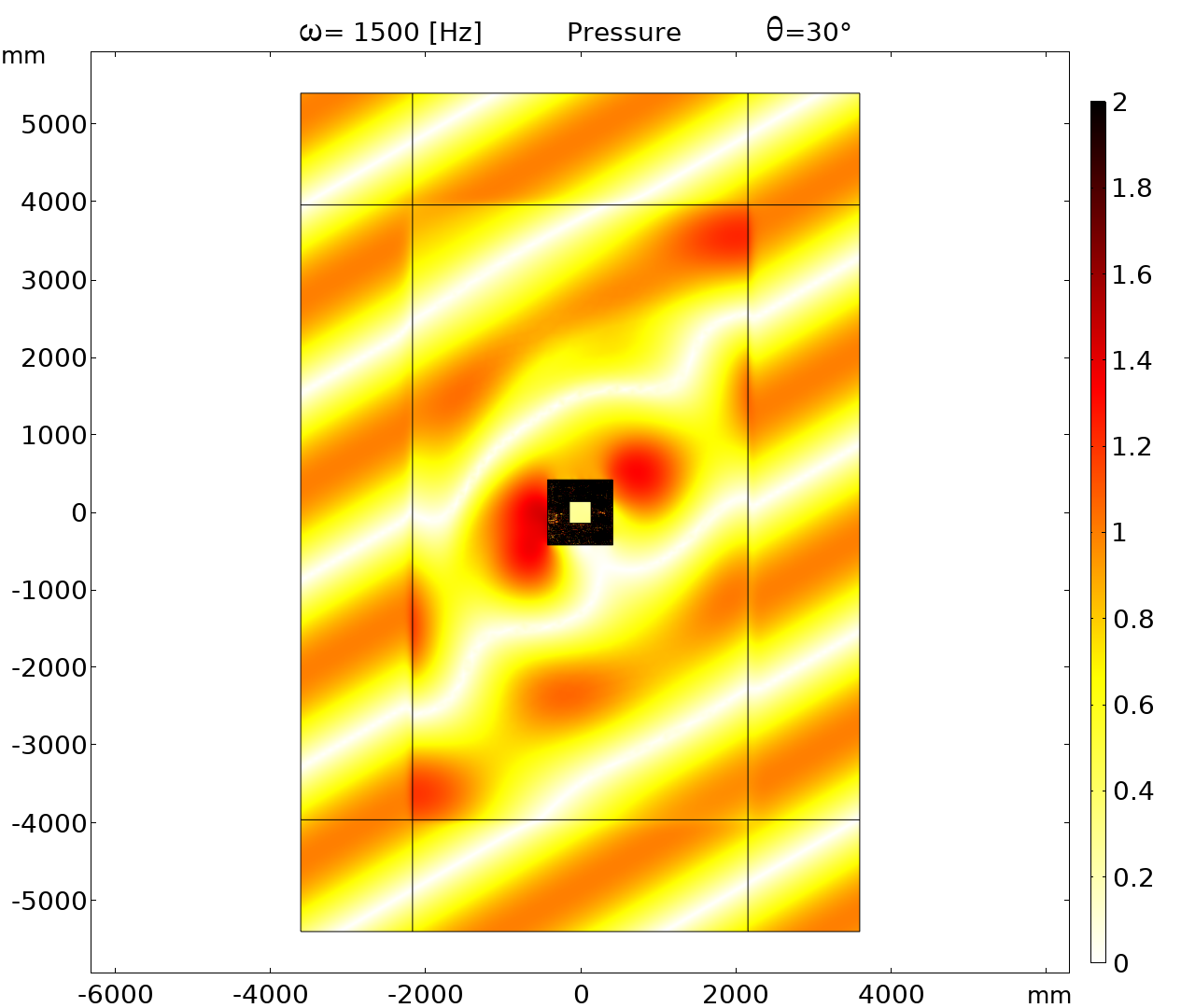}
	\end{minipage}
	\begin{minipage}[H]{0.32\textwidth}
		\includegraphics[width=\textwidth]{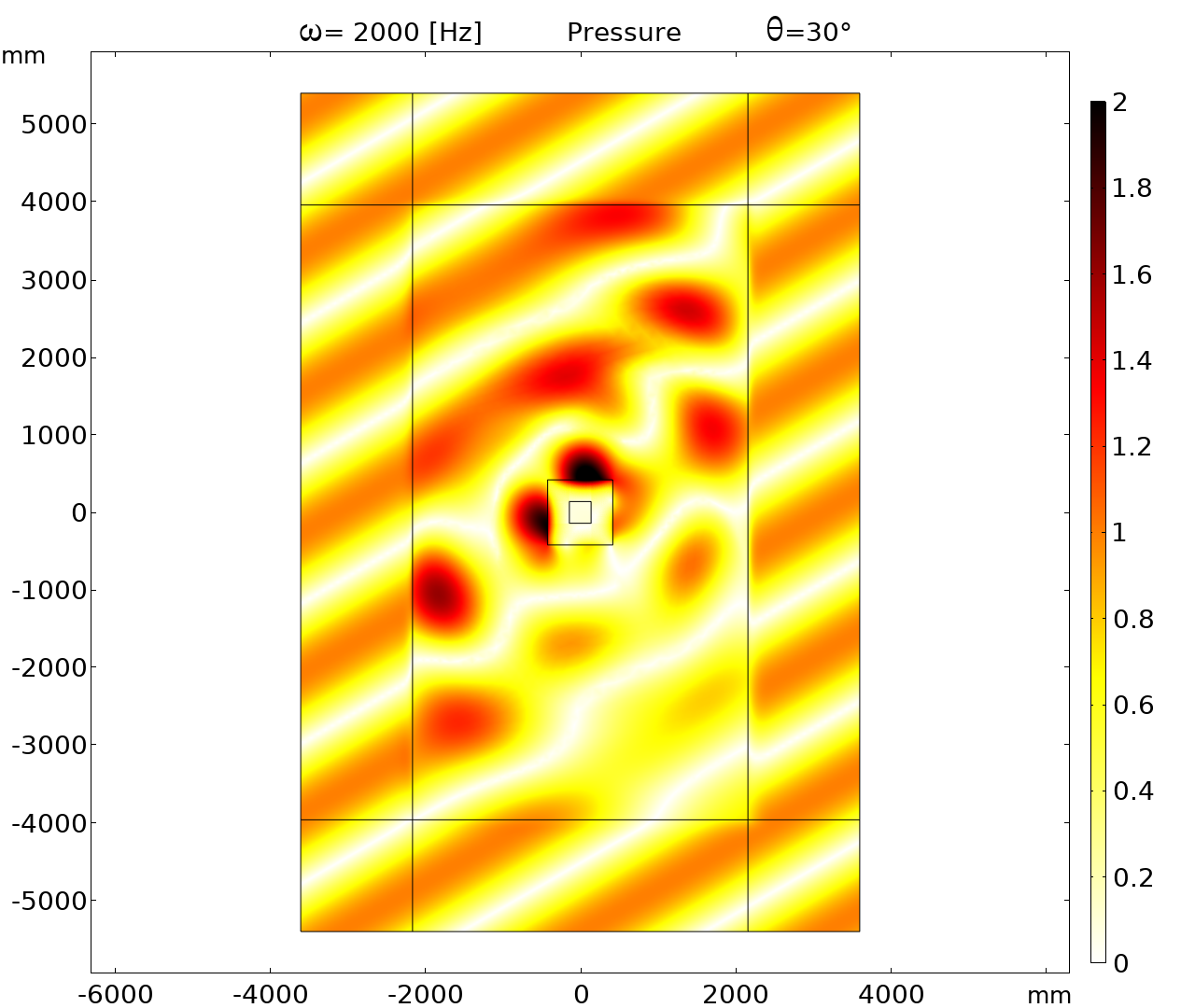}
	\end{minipage}
	\begin{minipage}[H]{0.32\textwidth}
		\includegraphics[width=\textwidth]{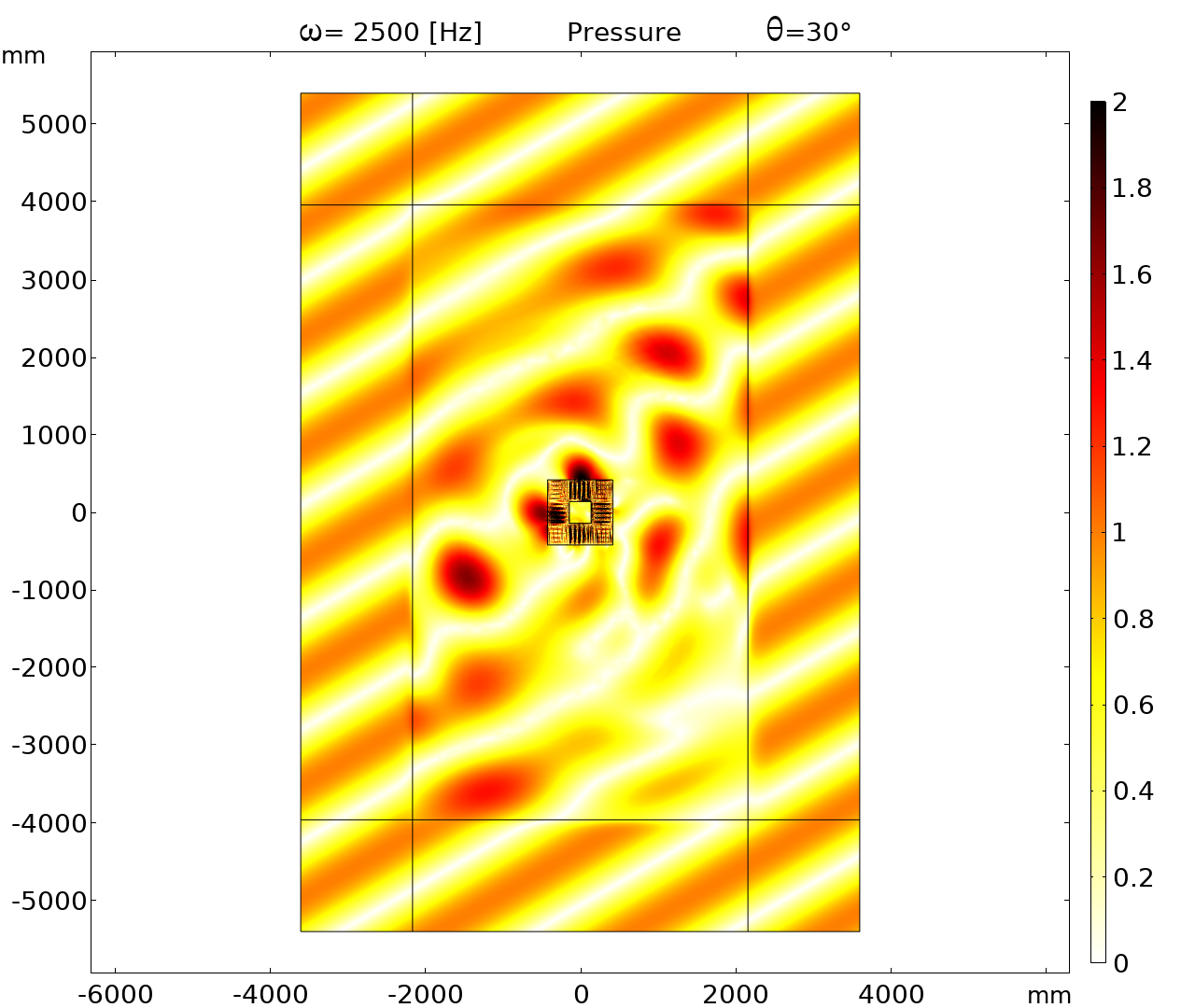}
	\end{minipage}
	\caption{
	Pressure incident wave's scattered field for a square made up of classic Cauchy material of side of 14 unit cells surrounded with a 14 unit cells thick metamaterial frame acting as a shield in the band-gap range.
	The angle of incidence is 30$^{\circ}$ and three frequencies are used.
	(\textit{top row}) Microstructured simulation.
	(\textit{bottom row}) Relaxed micromorphic simulation.
	(\textit{from left to right}) Frequency of 1.5 kHz (below the band-gap), 2 kHz (in the band-gap), and 2.5 kHz (above the band-gap).
	}
	\label{fig:Coll_compa_p_30}
\end{figure}

\begin{figure}[H]
	\centering
	\begin{minipage}[H]{0.32\textwidth}
		\includegraphics[width=\textwidth]{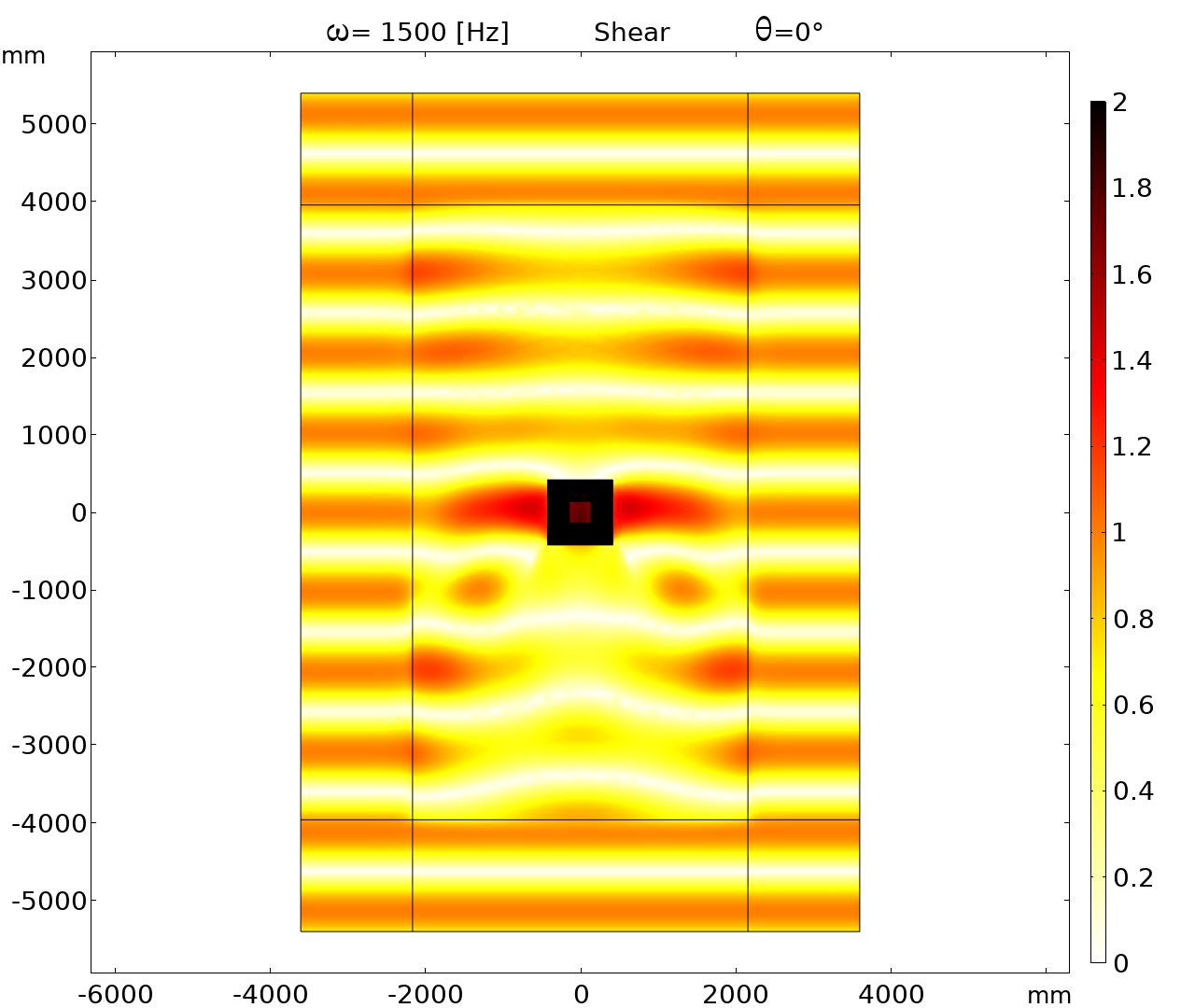}
	\end{minipage}
	\begin{minipage}[H]{0.32\textwidth}
		\includegraphics[width=\textwidth]{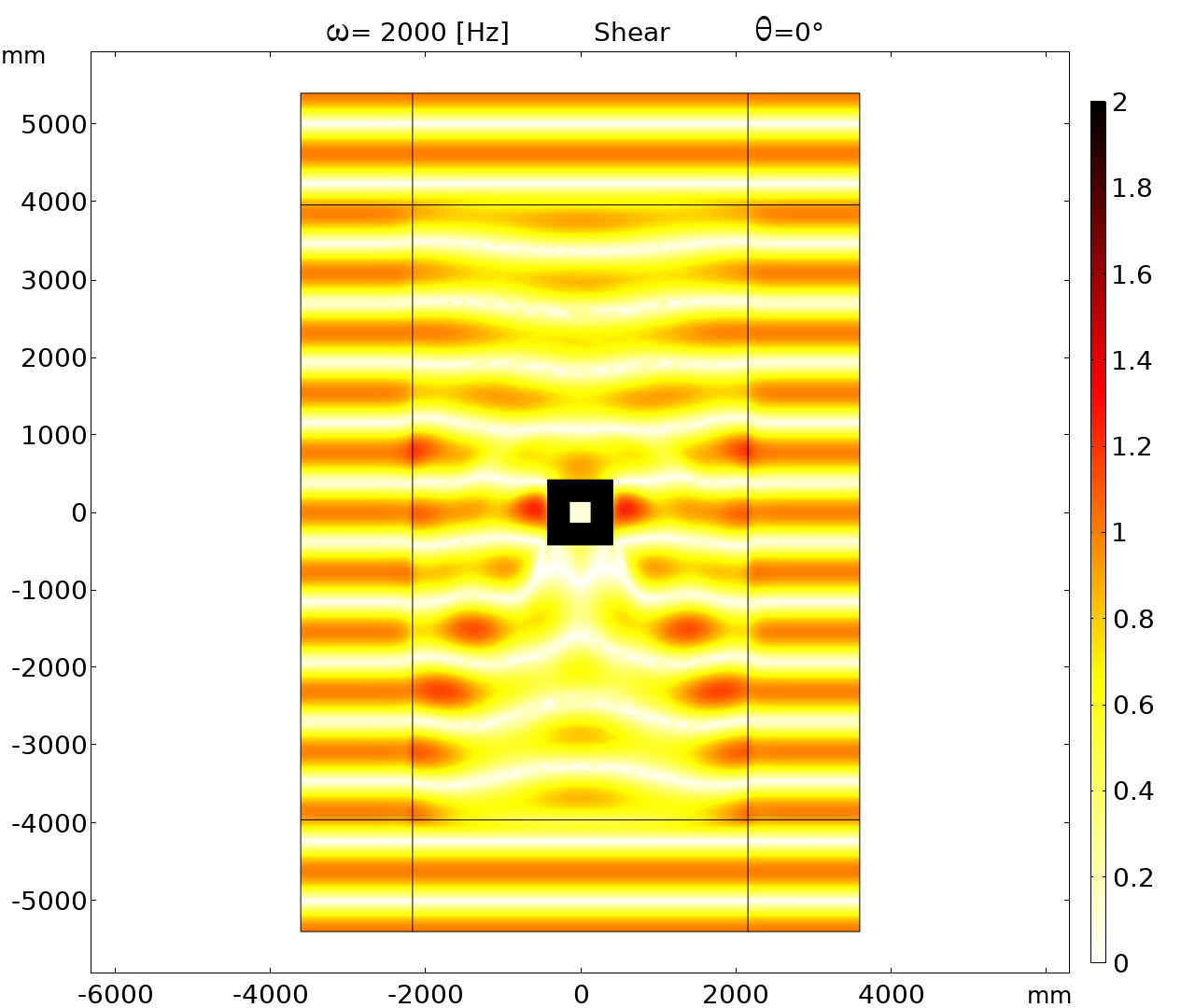}
	\end{minipage}
	\begin{minipage}[H]{0.32\textwidth}
		\includegraphics[width=\textwidth]{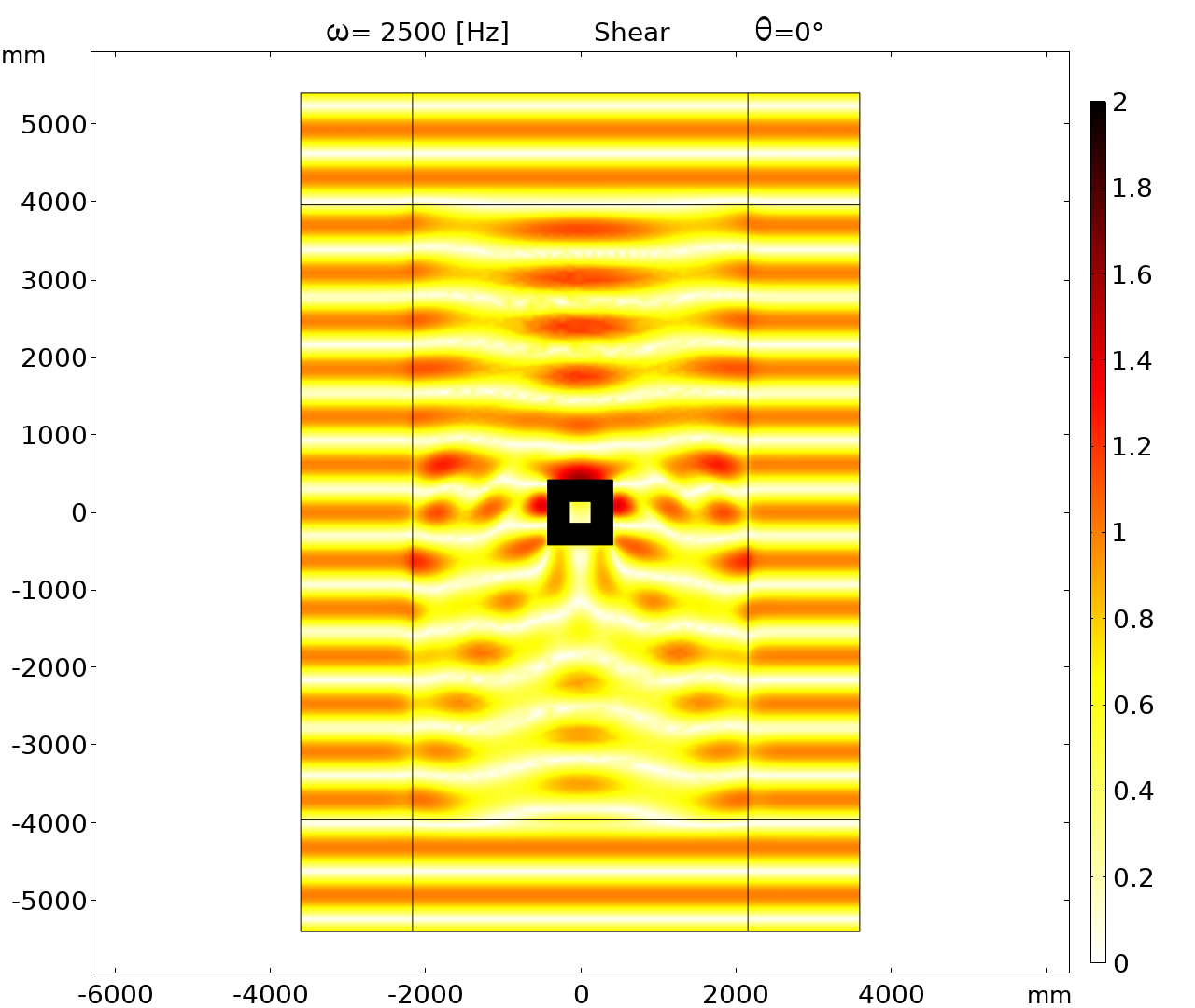}
	\end{minipage}
	\begin{minipage}[H]{0.32\textwidth}
		\includegraphics[width=\textwidth]{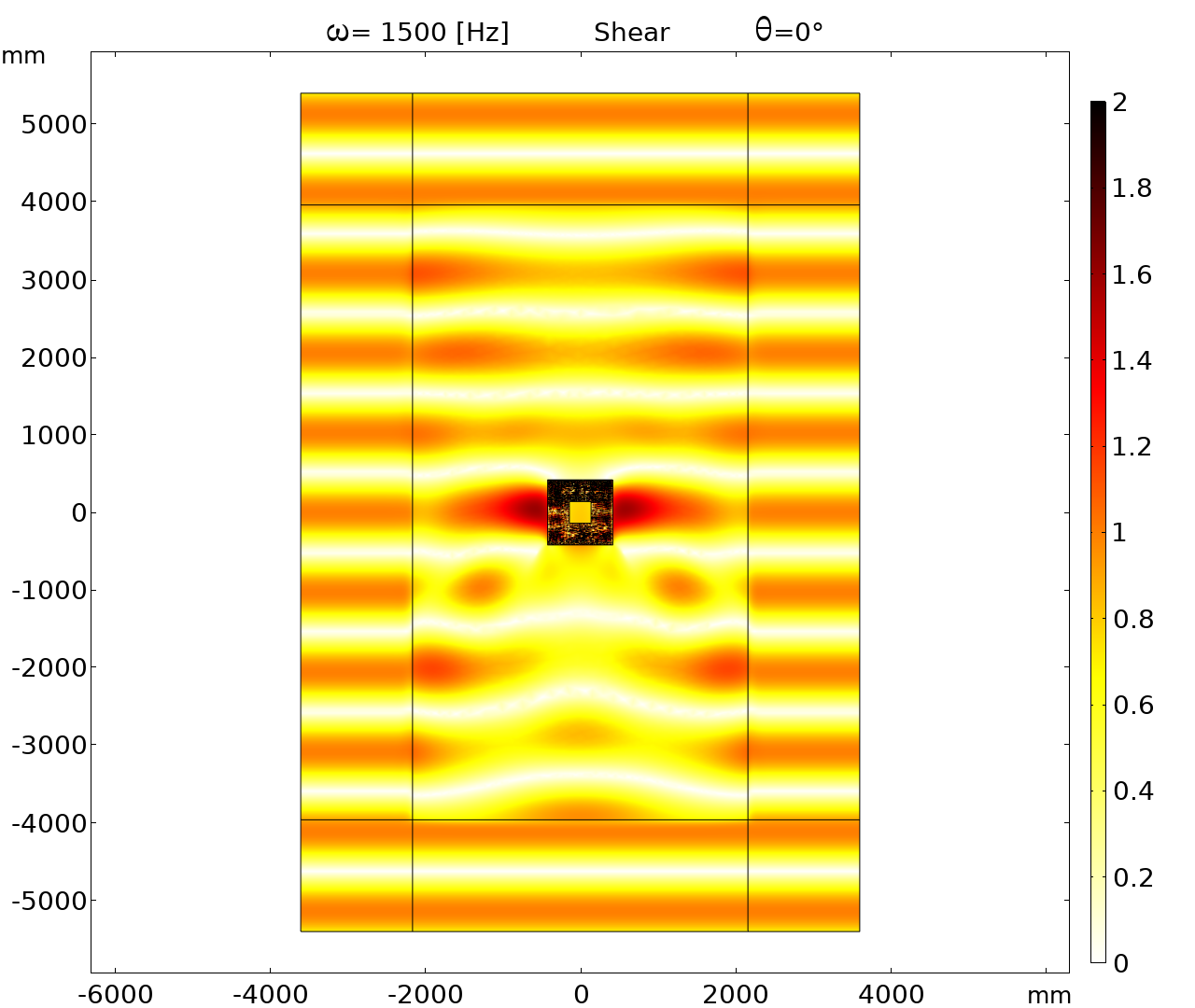}
	\end{minipage}
	\begin{minipage}[H]{0.32\textwidth}
		\includegraphics[width=\textwidth]{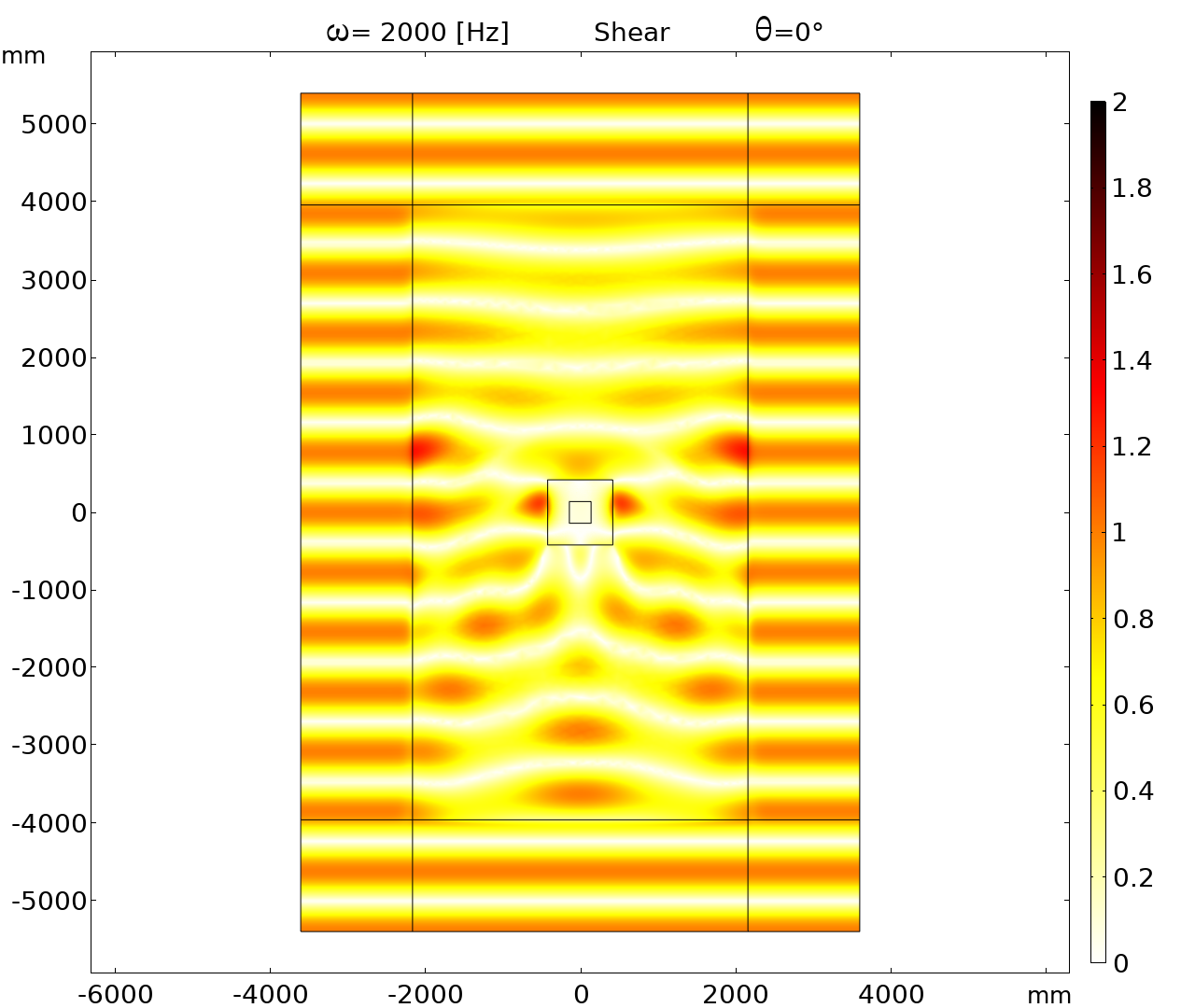}
	\end{minipage}
	\begin{minipage}[H]{0.32\textwidth}
		\includegraphics[width=\textwidth]{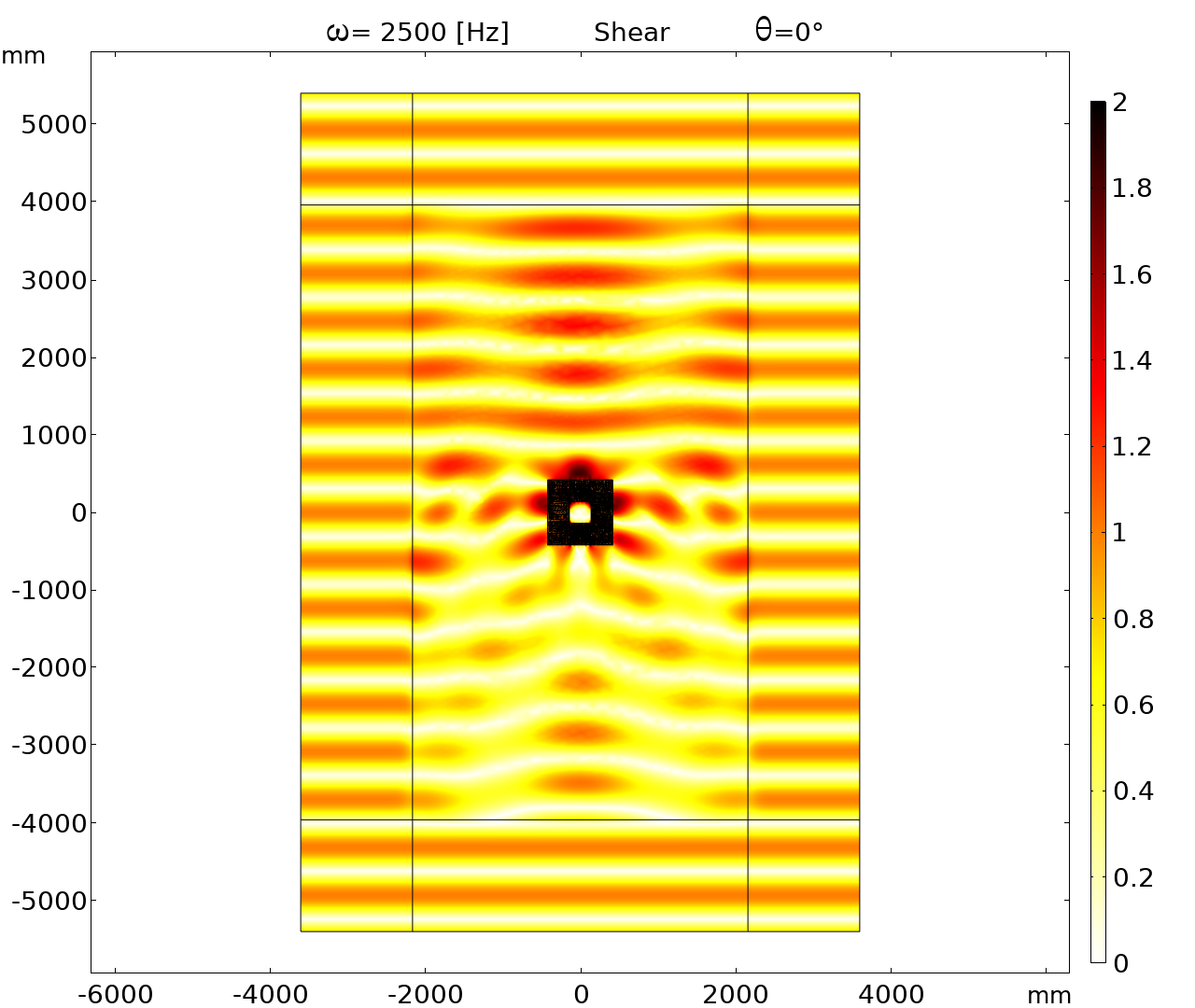}
	\end{minipage}
	\caption{
	Shear incident wave's scattered field for a square made up of classic Cauchy material of side of 14 unit cells surrounded with a 14 unit cells thick metamaterial frame acting as a shield in the band-gap range.
	The angle of incidence is 0$^{\circ}$ and three frequencies are used.
	(\textit{top row}) Microstructured simulation.
	(\textit{bottom row}) Relaxed micromorphic simulation.
	(\textit{from left to right}) Frequency of 1.5 kHz (below the band-gap), 2 kHz (in the band-gap), and 2.5 kHz (above the band-gap).
	}
	\label{fig:Coll_compa_s_00}
\end{figure}

\begin{figure}[H]
	\centering
	\begin{minipage}[H]{0.32\textwidth}
		\includegraphics[width=\textwidth]{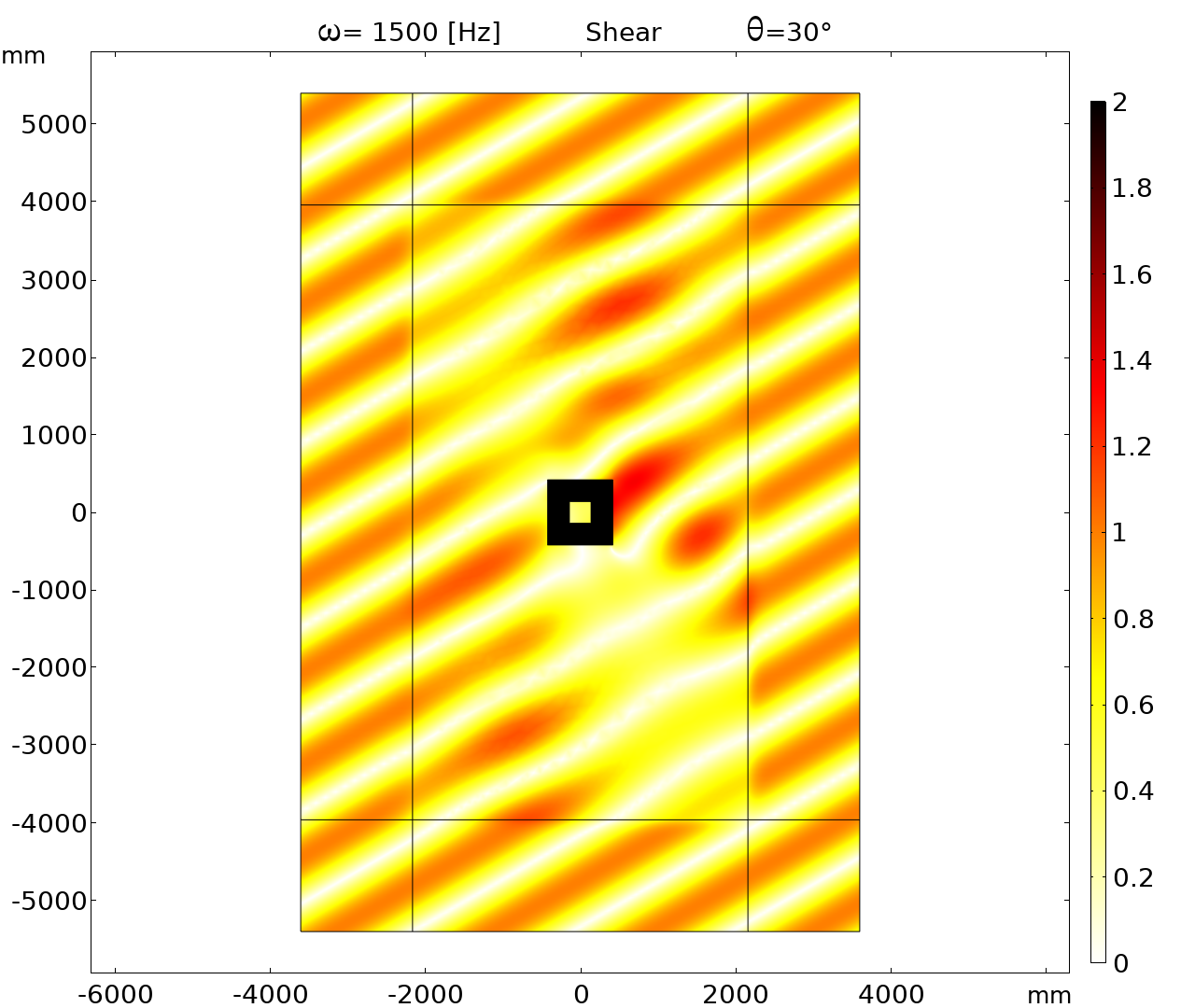}
	\end{minipage}
	\begin{minipage}[H]{0.32\textwidth}
		\includegraphics[width=\textwidth]{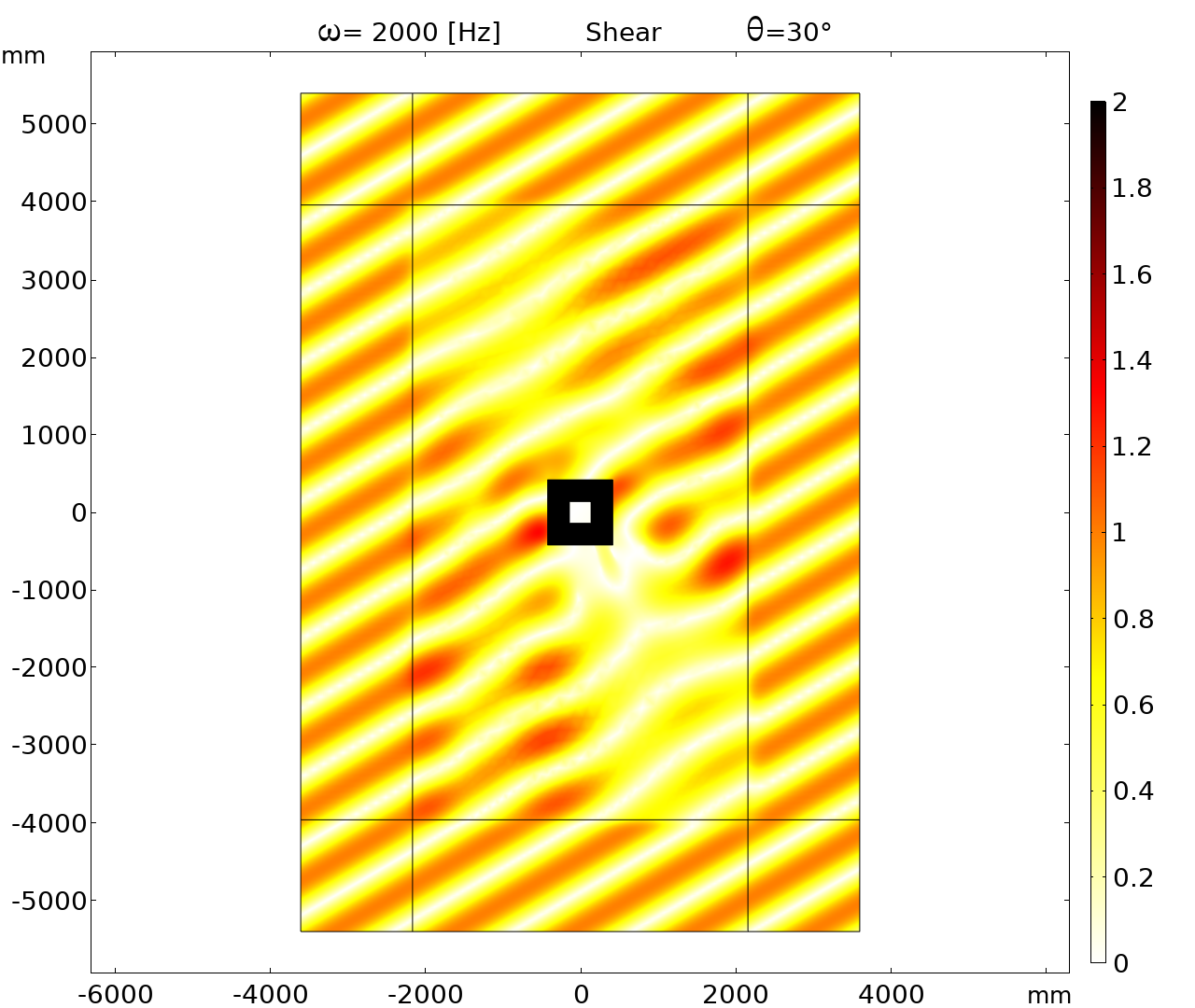}
	\end{minipage}
	\begin{minipage}[H]{0.32\textwidth}
		\includegraphics[width=\textwidth]{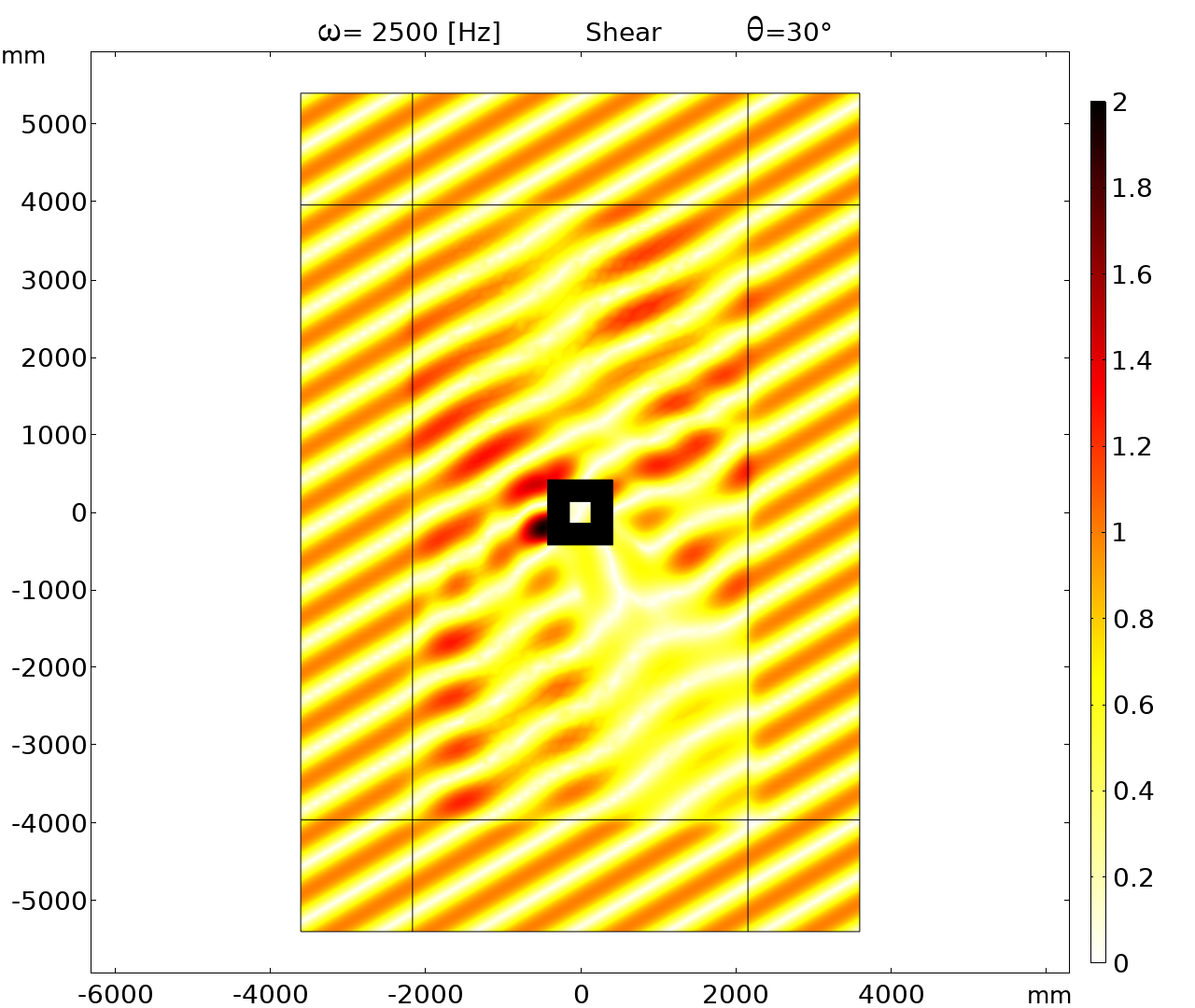}
	\end{minipage}
	\begin{minipage}[H]{0.32\textwidth}
		\includegraphics[width=\textwidth]{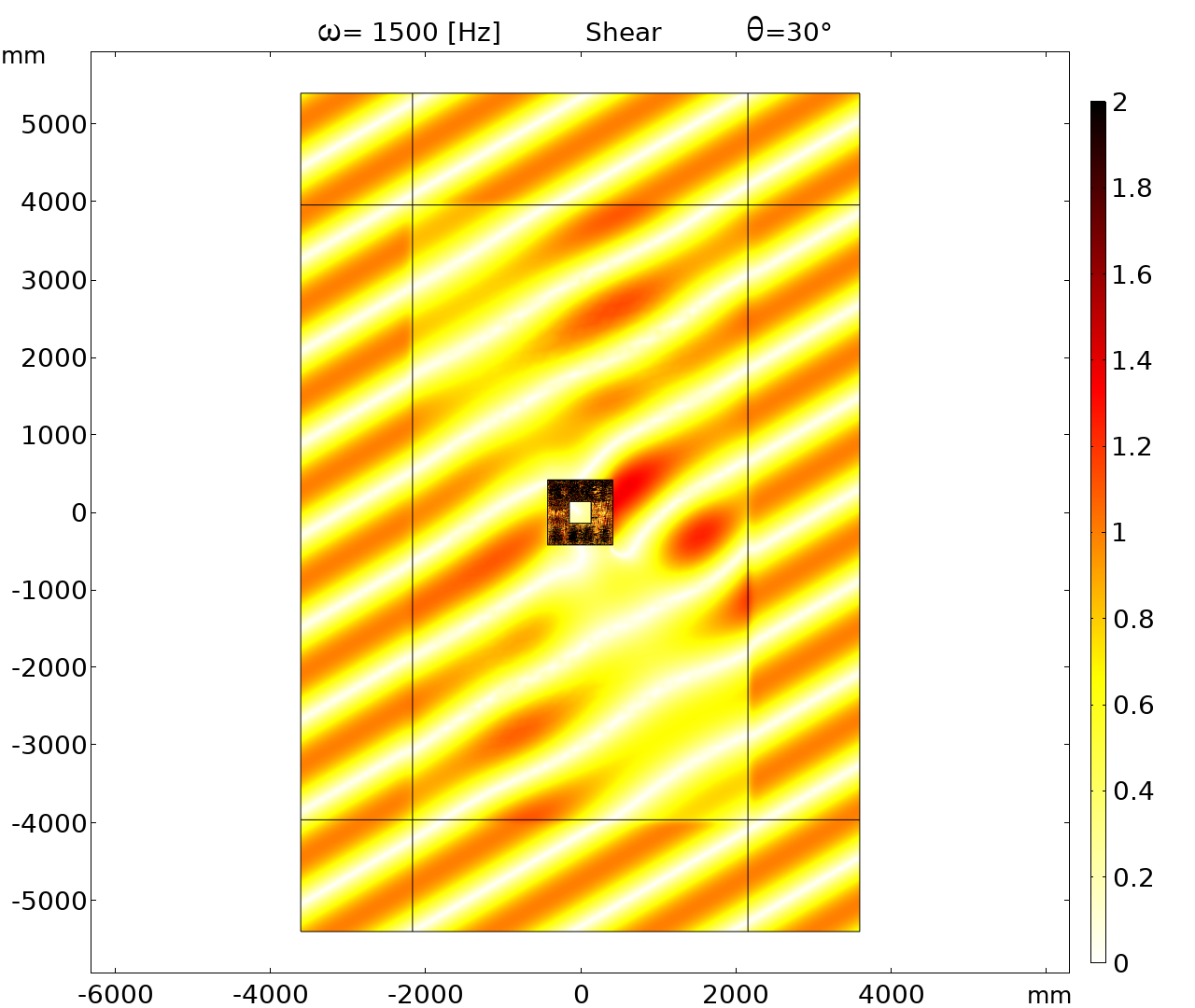}
	\end{minipage}
	\begin{minipage}[H]{0.32\textwidth}
		\includegraphics[width=\textwidth]{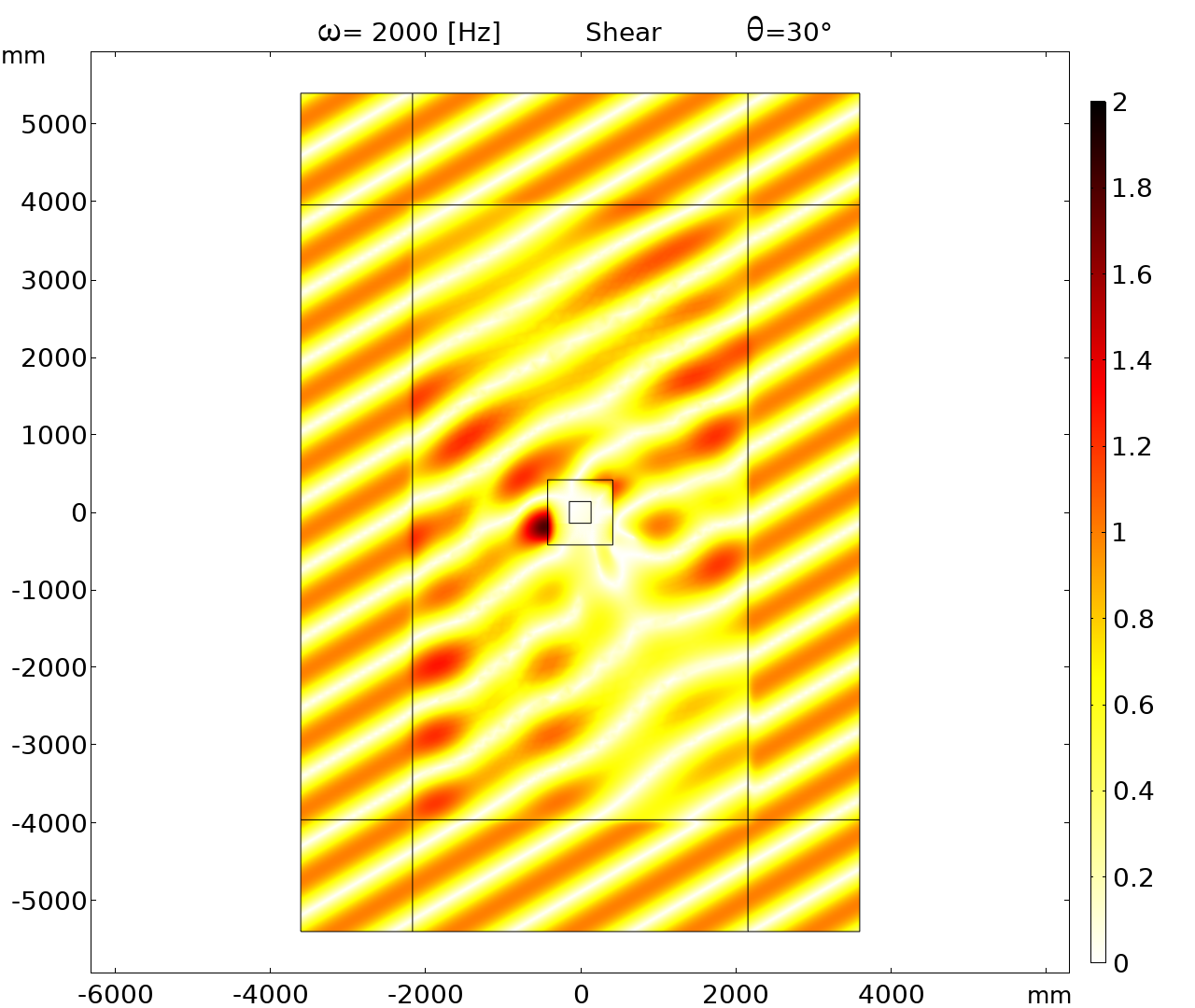}
	\end{minipage}
	\begin{minipage}[H]{0.32\textwidth}
		\includegraphics[width=\textwidth]{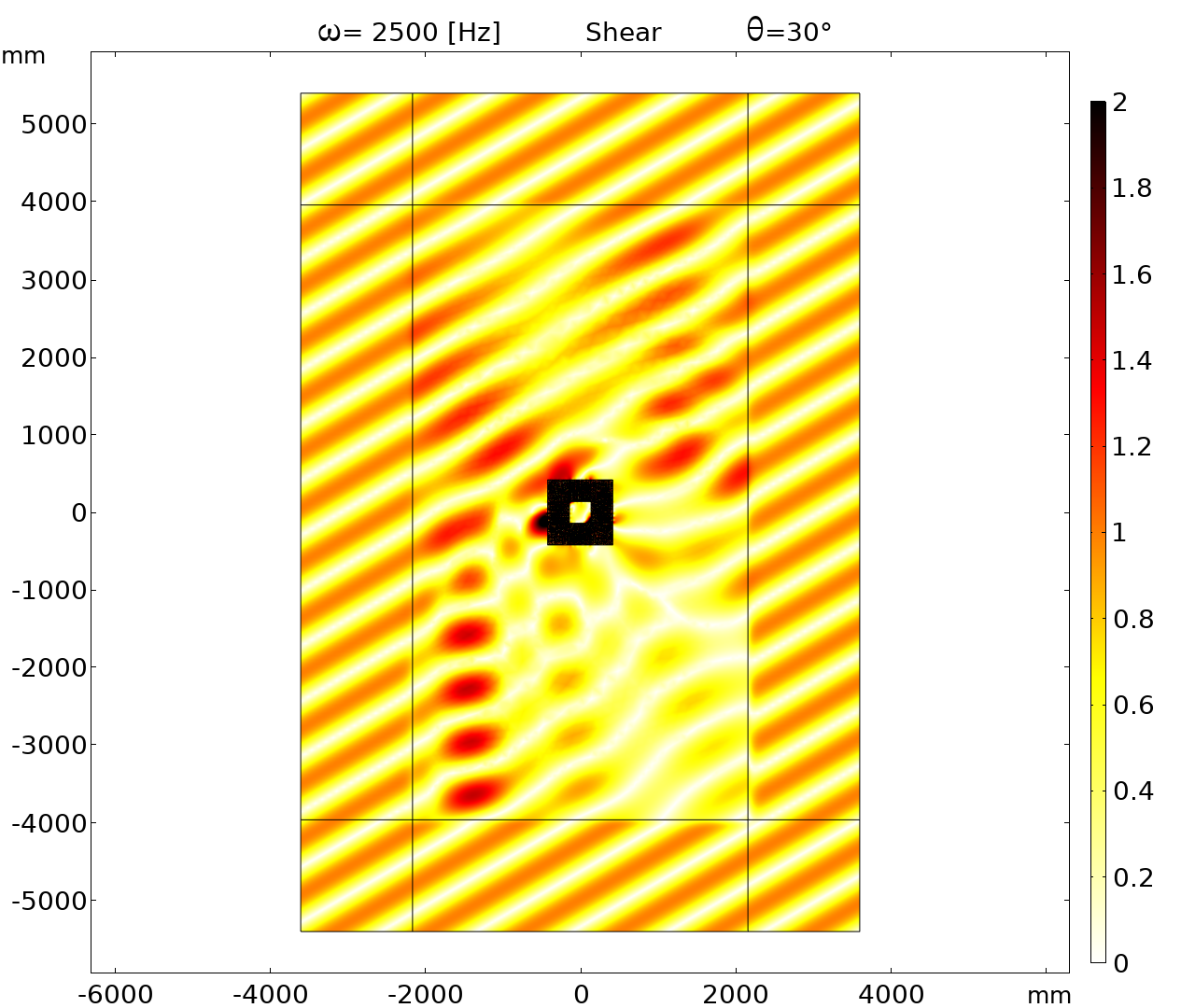}
	\end{minipage}
	\caption{
	Shear incident wave's scattered field for a square made up of classic Cauchy material of side of 14 unit cells surrounded with a 14 unit cells thick metamaterial frame acting as a shield in the band-gap range.
	The angle of incidence is 30$^{\circ}$ and three frequencies are used.
	(\textit{top row}) Microstructured simulation.
	(\textit{bottom row}) Relaxed micromorphic simulation.
	(\textit{from left to right}) Frequency of 1.5 kHz (below the band-gap), 2 kHz (in the band-gap), and 2.5 kHz (above the band-gap).
	}
	\label{fig:Coll_compa_s_30}
\end{figure}

In this subsection we start exploring the shield meta-structure for a metamaterial's sample size of 14 unit cells thick.
This size still allows a direct comparison with the detailed finite element simulation.
It is evident (see Fig.~\ref{fig:Coll_compa_p_00}--\ref{fig:Coll_compa_s_30}) that the relaxed micromorphic model gives excellent results for a wide range of frequencies (broadband) and for more than just the angles used to fit the parameters (anisotropy), and this for both pressure and shear waves.

\subsection{Exploring larger structures via the relaxed micromorphic material}
\label{sec:exploring}
\begin{figure}[H]
	\centering
	\begin{minipage}[H]{0.32\textwidth}
		\includegraphics[width=\textwidth]{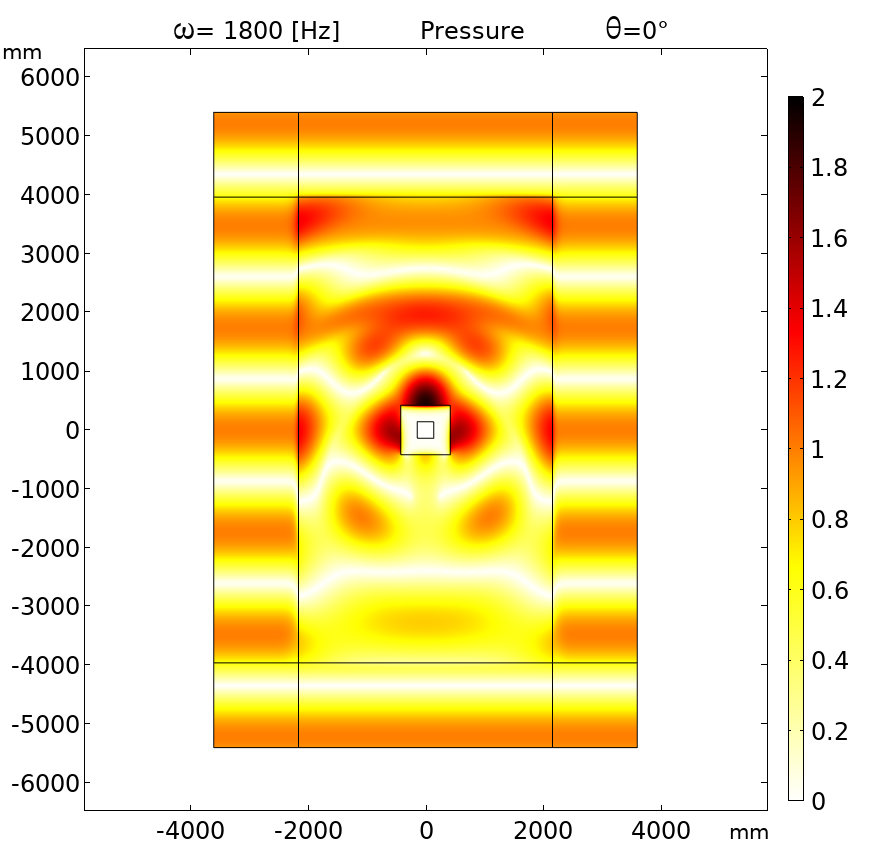}
	\end{minipage}
	\begin{minipage}[H]{0.32\textwidth}
		\includegraphics[width=\textwidth]{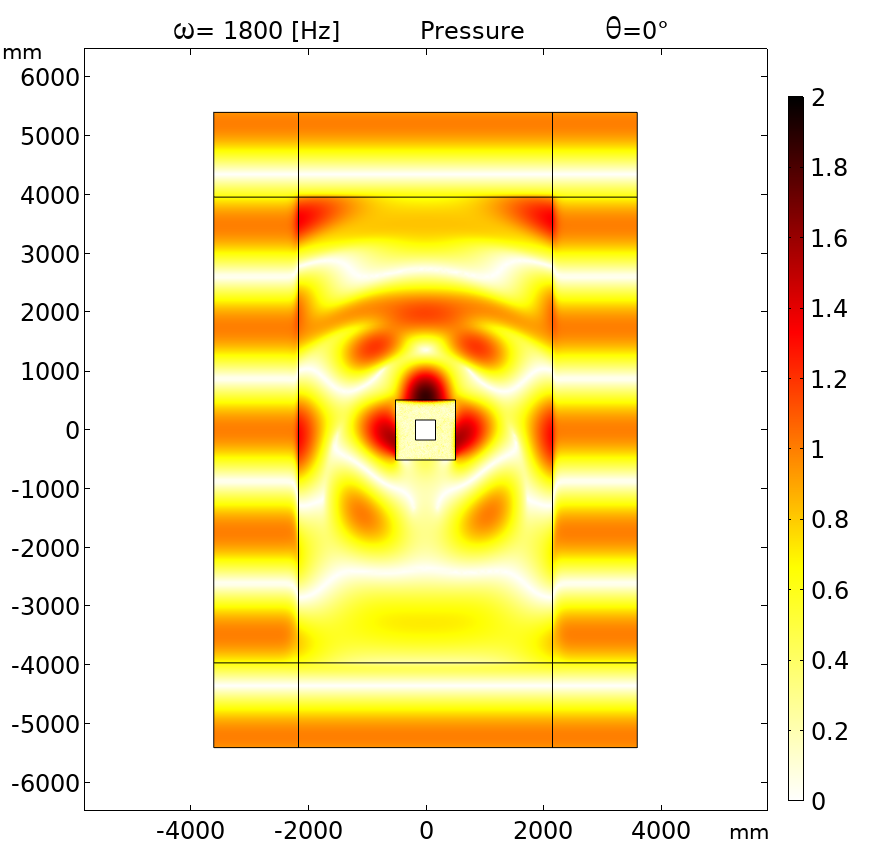}
	\end{minipage}
	\begin{minipage}[H]{0.32\textwidth}
		\includegraphics[width=\textwidth]{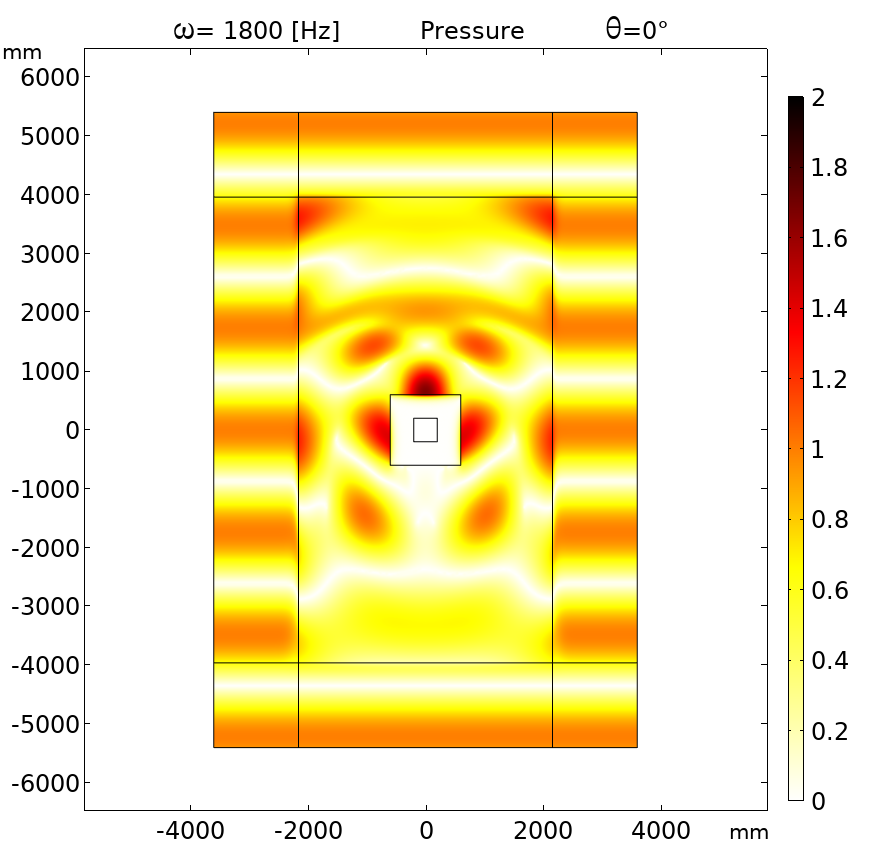}
	\end{minipage}
	\begin{minipage}[H]{0.32\textwidth}
		\includegraphics[width=\textwidth]{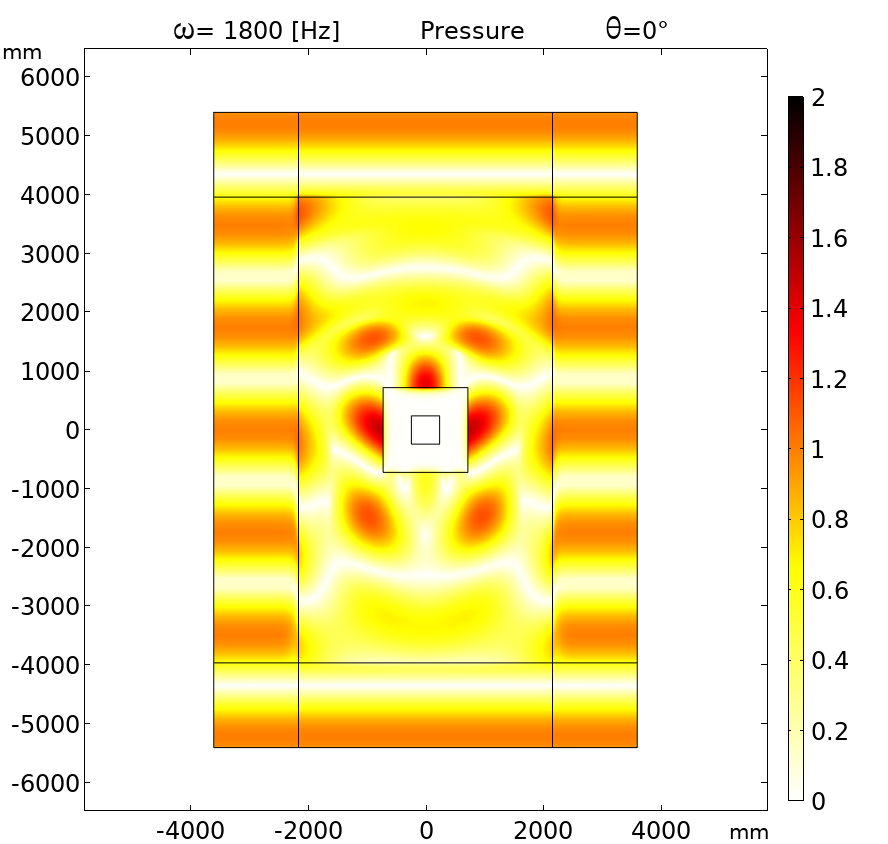}
	\end{minipage}
	\begin{minipage}[H]{0.32\textwidth}
		\includegraphics[width=\textwidth]{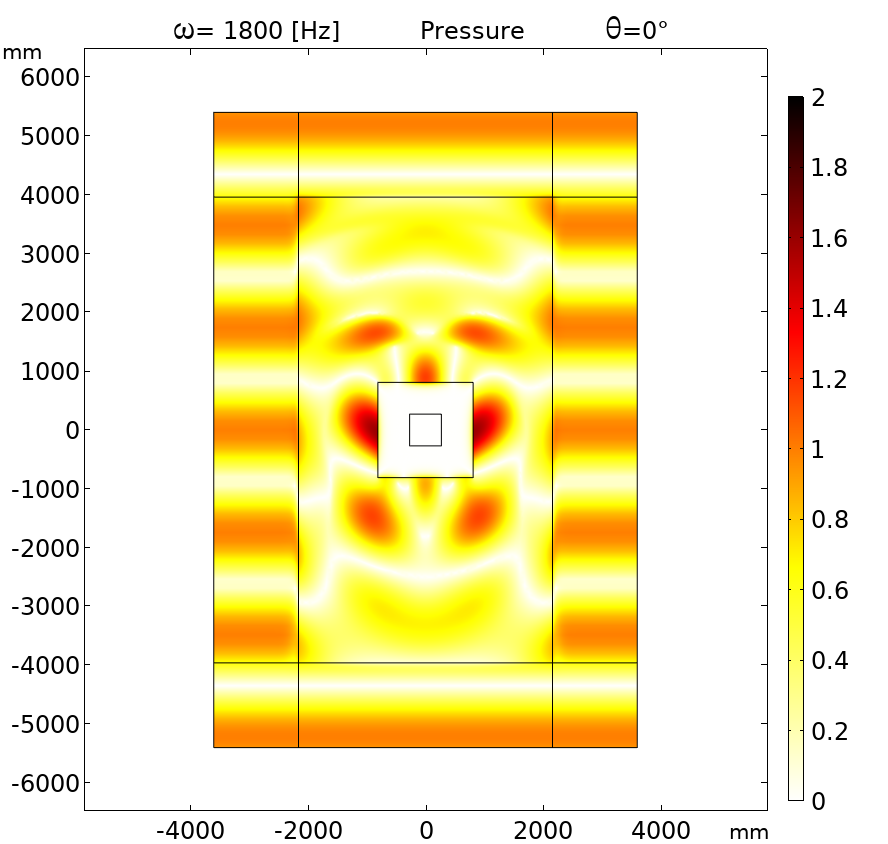}
	\end{minipage}
	\begin{minipage}[H]{0.32\textwidth}
		\includegraphics[width=\textwidth]{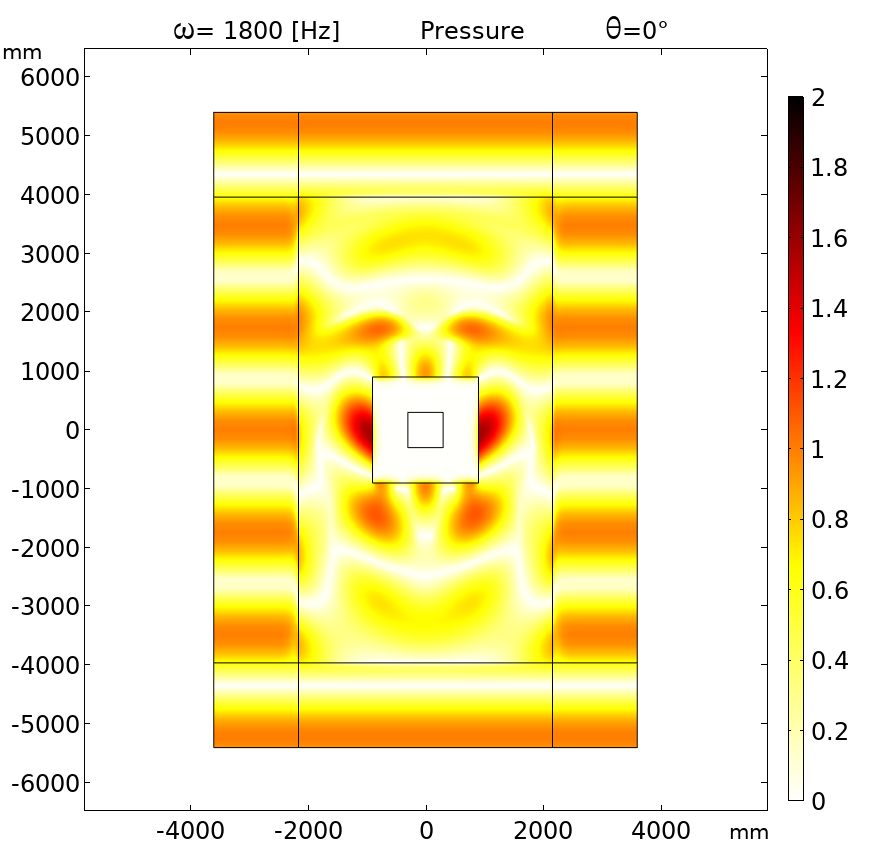}
	\end{minipage}
	\caption{
	Parametric study for a 1.8 kHz pressure wave and 0$^{\circ}$ angle of incidence.
	It is shown how the scattered field changes while changing the size of the portion of material shielded and the thickness of the metamaterial frame.
	(\textit{From left to right and top to bottom}) the thickness of the micro-structured material frame is equal to the side of the shielded square which is 14, 17, 20, 24, 27, 30 cells thick, respectively.
	}
	\label{fig:Coll_para_study_p_0_18E2}
\end{figure}

\begin{figure}[H]
	\centering
	\begin{minipage}[H]{0.32\textwidth}
		\includegraphics[width=\textwidth]{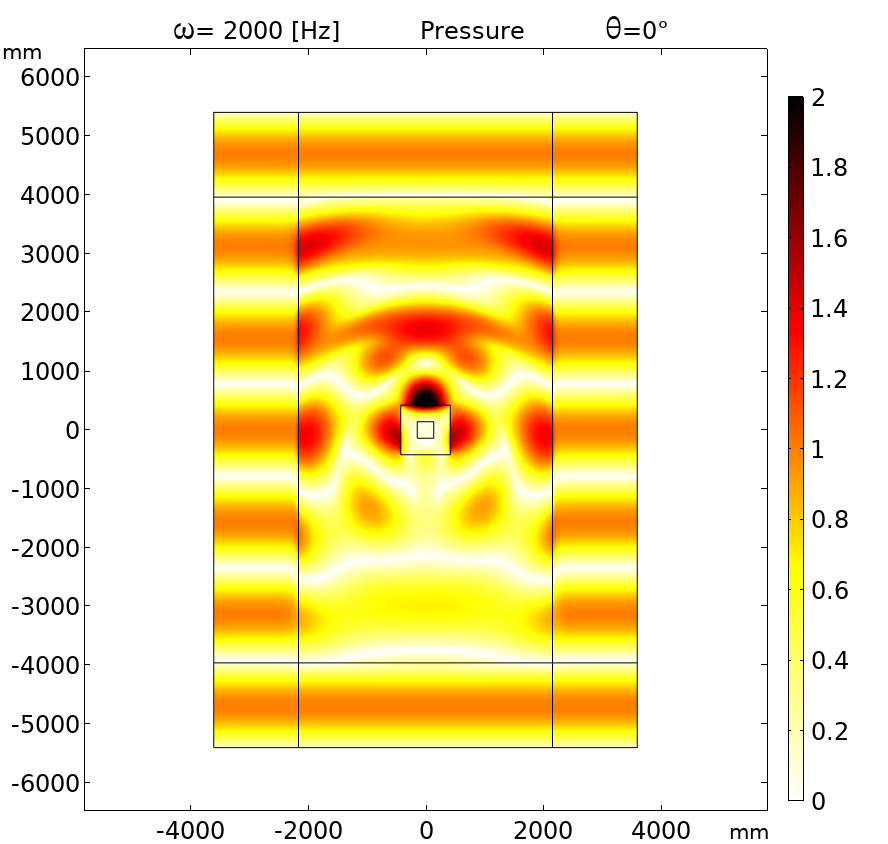}
	\end{minipage}
	\begin{minipage}[H]{0.32\textwidth}
		\includegraphics[width=\textwidth]{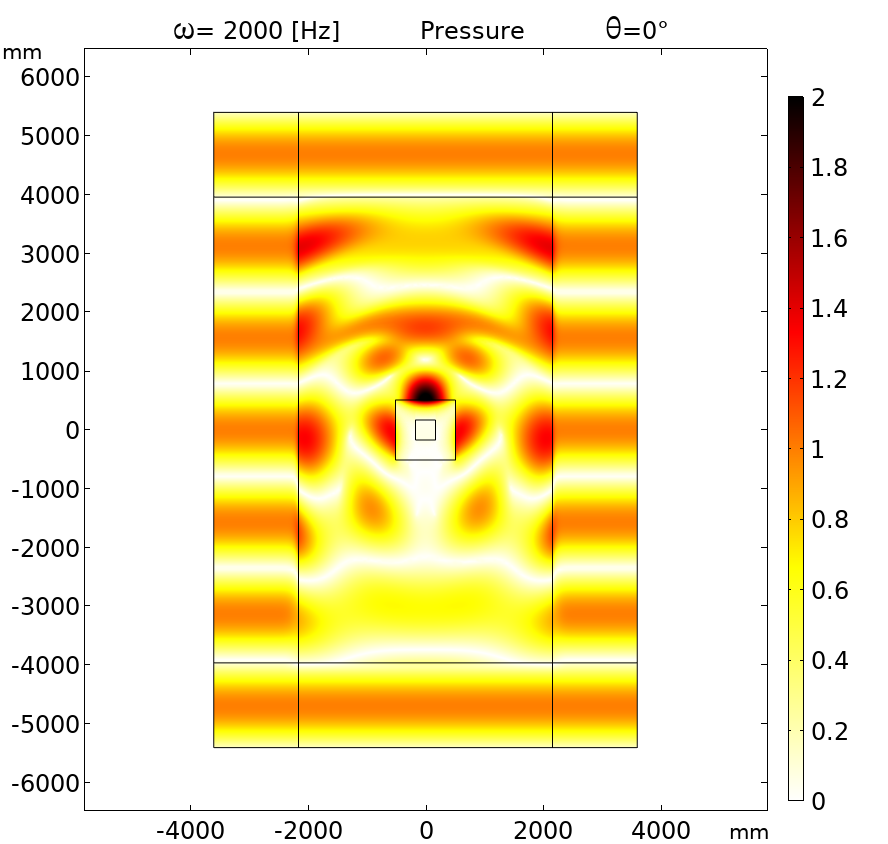}
	\end{minipage}
	\begin{minipage}[H]{0.32\textwidth}
		\includegraphics[width=\textwidth]{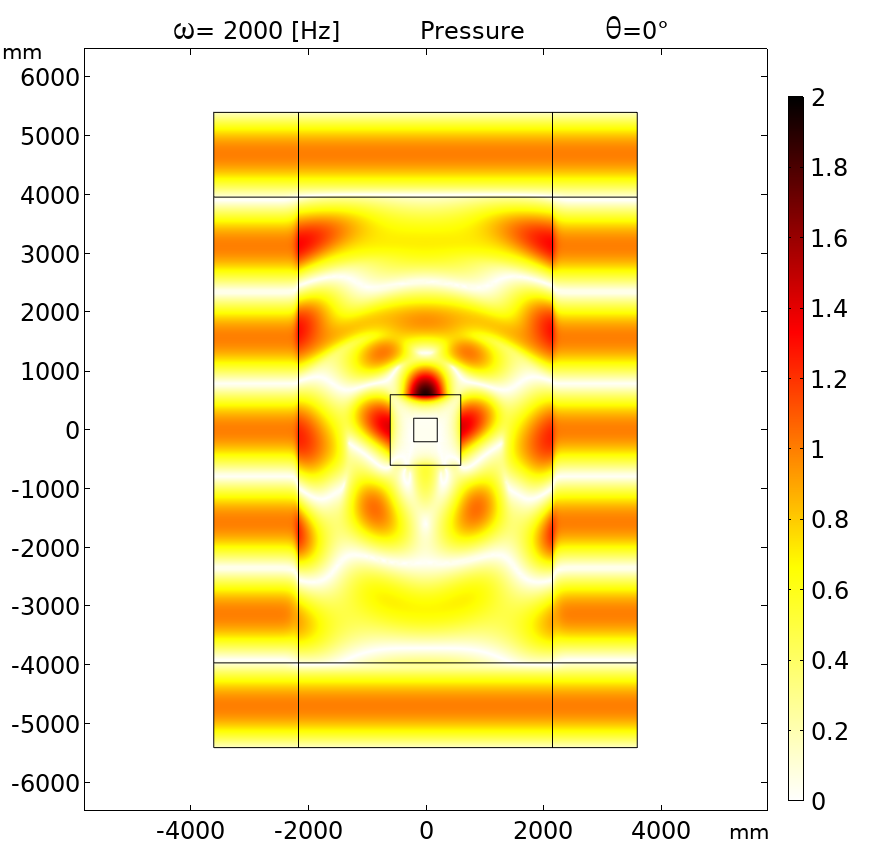}
	\end{minipage}
	\begin{minipage}[H]{0.32\textwidth}
		\includegraphics[width=\textwidth]{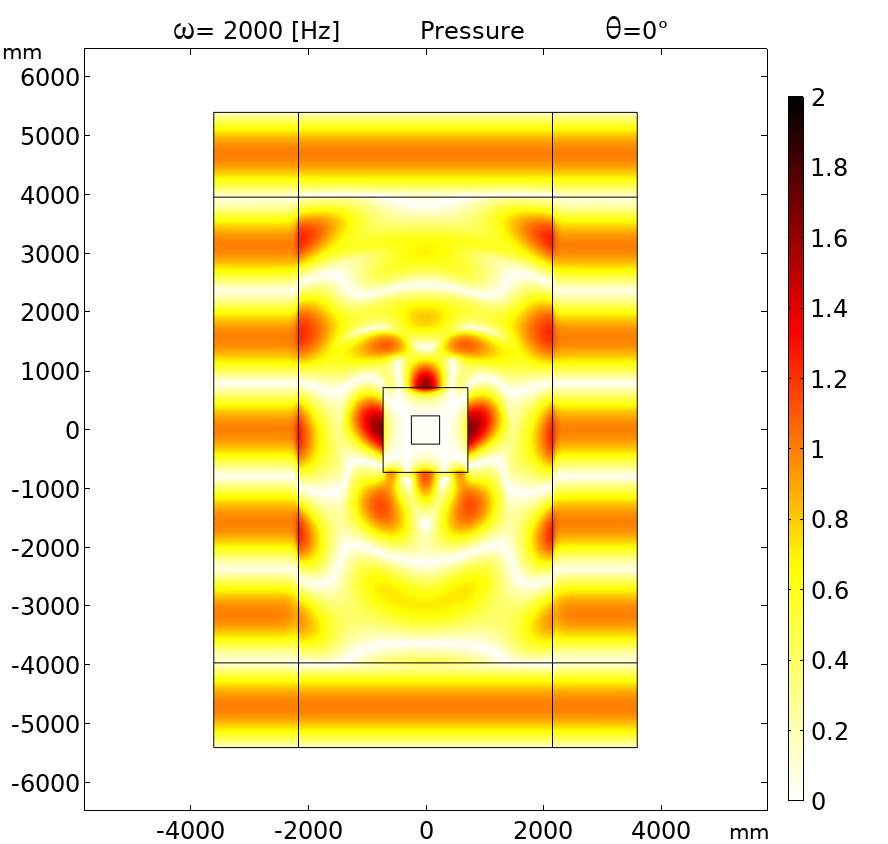}
	\end{minipage}
	\begin{minipage}[H]{0.32\textwidth}
		\includegraphics[width=\textwidth]{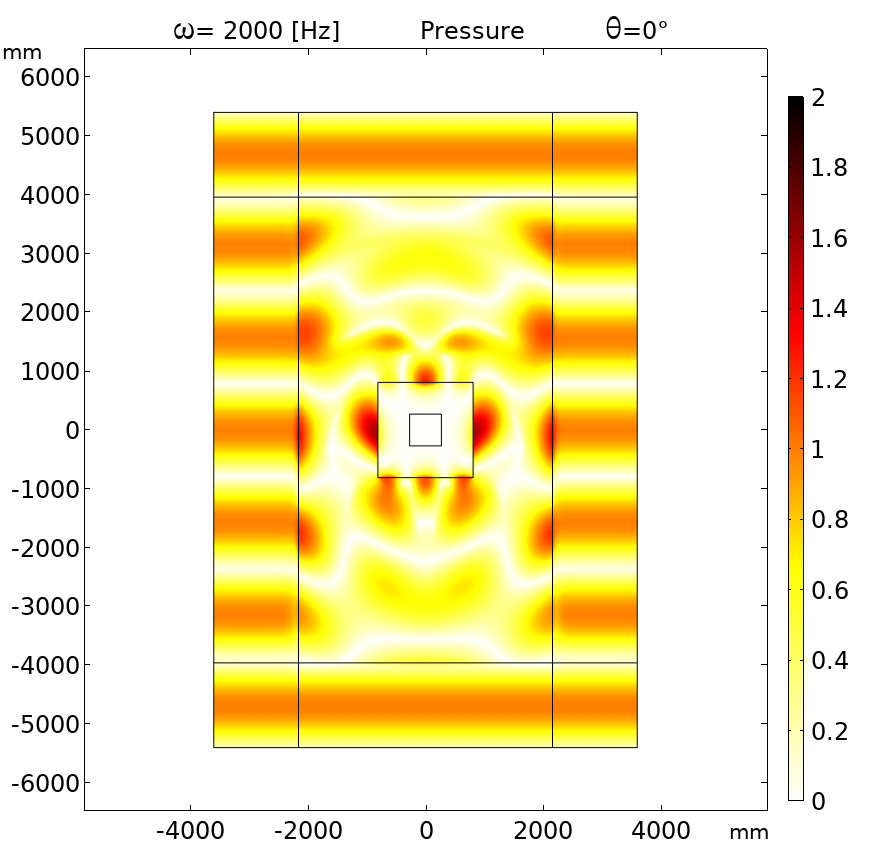}
	\end{minipage}
	\begin{minipage}[H]{0.32\textwidth}
		\includegraphics[width=\textwidth]{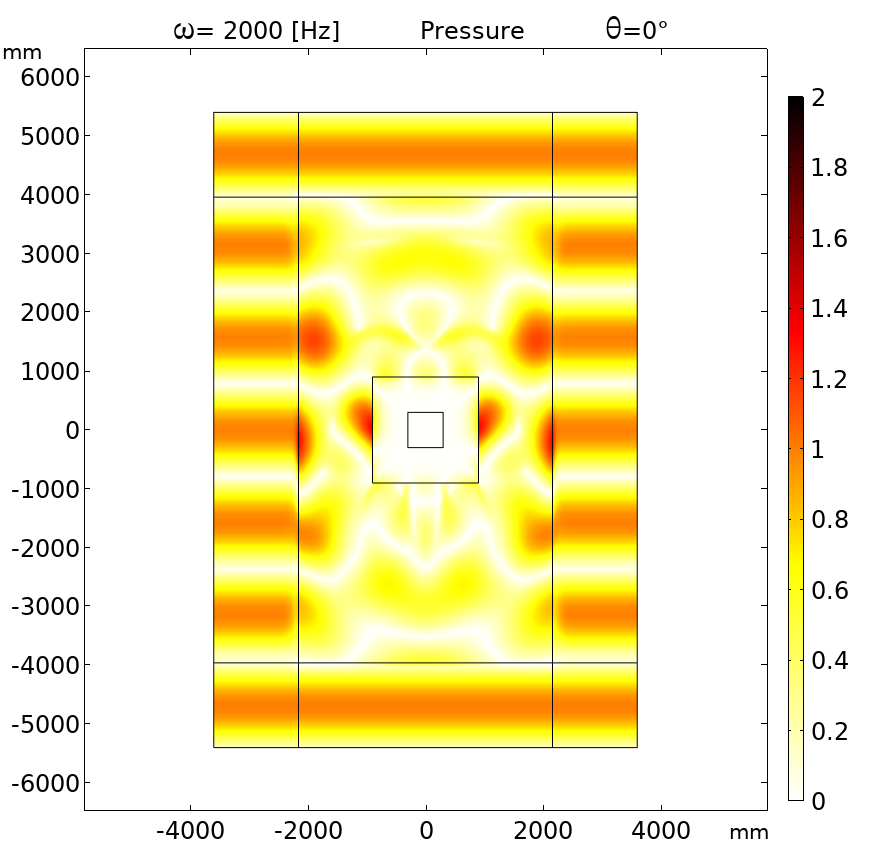}
	\end{minipage}
	\caption{
	Parametric study for a 2 kHz pressure wave and 0$^{\circ}$ angle of incidence.
	It is shown how the scattered field changes while changing the size of the portion of material shielded and the thickness of the metamaterial frame.
	(\textit{From left to right and top to bottom}) the thickness of the equivalent relaxed micromorphic material frame is equal to the side of the shielded square which is 14, 17, 20, 24, 27, 30 cells thick, respectively.
	}
	\label{fig:Coll_para_study_p_0_20E2}
\end{figure}

In this subsection we explore the meta-structures's scattering behaviours for increasingly bigger metamaterials' samples.
This study would have not been possible via detailed numerical simulations of real structures due to the non-linearly increasing computational time.
For the sake of brevity, we only present results for pressure incident waves, since the ones for the shear waves are completely analogous.
It is possible to infer from Fig.~\ref{fig:Coll_para_study_p_0_18E2} and Fig.~\ref{fig:Coll_para_study_p_0_20E2} that the scattered field changes radically while changing the size of the shield.
This fact is of great importance to optimize shielding devices in terms of the scattered field: while it is quite simple to design a metamaterial shield, it is not possible nowadays to optimize the reflected wave, but here, a calibration of the size of the shield gives the possibility of exploring the reflected energy and/or to focus energy on the metamaterial's boundary or in the near surrounding of the shield.
This, in turn, makes possible to focus energy in specific points for eventual subsequent re-use.

\section{Conclusions}
\label{sec:conclusion}
In this paper, we show that a relaxed micromorphic modeling of metamaterials can unveil their dynamic response for a wide range of frequencies and angles of propagation of the incident wave while using a limited number of constant (frequency- and angle- independent) parameters.
The versatility of the proposed model allows to explore large scale meta-structures to an extent that would not be otherwise feasible.
In turn, this opens new perspectives for the optimization of meta-structures both in terms of shielding capacity and of re-use of scattered energy.
In future works, we will use the results obtained in this paper as a stepping stone towards the conception of sustainable meta-structures combining together different metamaterials as well as classical homogeneous materials with the aim of controlling elastic waves while recovering energy.

{\scriptsize
	\paragraph{{\scriptsize Acknowledgements.}}
	Angela Madeo and Gianluca Rizzi acknowledges support from the European Commission through the funding of the ERC Consolidator Grant META-LEGO, N° 101001759.
	Angela Madeo and Gianluca Rizzi acknowledge funding from the French Research Agency ANR, “METASMART” (ANR-17CE08-0006).
	Angela Madeo thanks IUF (Institut Universitaire de France) for its support.
	Patrizio Neff acknowledges support in the framework of the DFG-Priority Programme 2256 ``Variational Methods for Predicting Complex Phenomena in Engineering Structures and Materials'', Neff 902/10-1, Project-No. 440935806.
}


\begingroup
\setstretch{0.8}
\setlength\bibitemsep{3pt}
\printbibliography

@article{brun2012vortex,
  title={Vortex-type elastic structured media and dynamic shielding},
  author={Brun, Michele and Jones, Ian S and Movchan, Alexander B},
  journal={Proceedings of the Royal Society A: Mathematical, Physical and Engineering Sciences},
  volume={468},
  number={2146},
  pages={3027--3046},
  year={2012},
  publisher={The Royal Society Publishing}
}

@article{la2017conception,
  title={Conception of a 3D metamaterial-based foundation for static and seismic protection of fuel storage tanks},
  author={La Salandra, Vincenzo and Wenzel, Moritz and Bursi, Oreste S and Carta, Giorgio and Movchan, Alexander B},
  journal={Frontiers in Materials},
  volume={4},
  pages={30},
  year={2017},
  publisher={Frontiers}
}

@article{miniaci2016spider,
  title={Spider web-inspired acoustic metamaterials},
  author={Miniaci, Marco and Krushynska, Anastasiia and Movchan, Alexander B and Bosia, Federico and Pugno, Nicola M},
  journal={Applied Physics Letters},
  volume={109},
  number={7},
  pages={071905},
  year={2016},
  publisher={AIP Publishing LLC}
}

@article{bordiga2019prestress,
  title={Prestress tuning of negative refraction and wave channeling from flexural sources},
  author={Bordiga, Giovanni and Cabras, Luigi and Piccolroaz, Andrea and Bigoni, Davide},
  journal={Applied Physics Letters},
  volume={114},
  number={4},
  pages={041901},
  year={2019},
  publisher={AIP Publishing LLC}
}

@article{bacigalupo2020design,
  title={Design of tunable acoustic metamaterials with periodic piezoelectric microstructure},
  author={Bacigalupo, Andrea and De Bellis, Maria Laura and Misseroni, Diego},
  journal={Extreme Mechanics Letters},
  volume={40},
  pages={100977},
  year={2020},
  publisher={Elsevier}
}

@article{barnwell2016antiplane,
  title={Antiplane elastic wave propagation in pre-stressed periodic structures; tuning, band gap switching and invariance},
  author={Barnwell, Ellis G and Parnell, William J and Abrahams, I David},
  journal={Wave Motion},
  volume={63},
  pages={98--110},
  year={2016},
  publisher={Elsevier}
}

@article{ragonese2021prediction,
  title={Prediction of local resonance band gaps in 2D elastic metamaterials via Bloch mode identification},
  author={Ragonese, A and Nouh, M},
  journal={Wave Motion},
  pages={102734},
  year={2021},
  publisher={Elsevier}
}

@article{dagostino_effective_2020,
	title = {Effective {description} of {anisotropic} {wave} {dispersion} in {mechanical} {band}-{gap} {metamaterials} via the {relaxed} {micromorphic} {model}},
	volume = {139},
	issn = {0374-3535, 1573-2681},
	url = {http://link.springer.com/10.1007/s10659-019-09753-9},
	doi = {10.1007/s10659-019-09753-9},
	language = {en},
	number = {2},
	urldate = {2021-04-23},
	journal = {Journal of Elasticity},
	author = {d’Agostino, Marco Valerio and Barbagallo, Gabriele and Ghiba, Ionel-Dumitrel and Eidel, Bernhard and Neff, Patrizio and Madeo, Angela},
	month = may,
	year = {2020},
	pages = {299--329},
}

@article{rizzi_exploring_2021,
	title = {Exploring {metamaterials}’ {structures} {through} the {relaxed} {micromorphic} {model}: {switching} an {acoustic} {screen} {into} an {acoustic} {absorber}},
	volume = {7},
	issn = {2296-8016},
	shorttitle = {Exploring {Metamaterials}’ {Structures} {Through} the {Relaxed} {Micromorphic} {Model}},
	url = {https://www.frontiersin.org/articles/10.3389/fmats.2020.589701/full},
	doi = {10.3389/fmats.2020.589701},
	urldate = {2021-04-23},
	journal = {Frontiers in Materials},
	author = {Rizzi, Gianluca and Collet, Manuel and Demore, Félix and Eidel, Bernhard and Neff, Patrizio and Madeo, Angela},
	month = mar,
	year = {2021},
	pages = {589701},
}

@article{rizzi2020towards,
  title={Towards the conception of complex engineering meta-structures: relaxed-micromorphic modelling of mechanical diodes},
  author={Rizzi, Gianluca and Tallarico, Domenico and Neff, Patrizio and Madeo, Angela},
  journal={(arXiv:2012.11192) to appear in Wave Motion},
  year={2021}
}

@article{rizzi2021boundary,
  title={Boundary and interface conditions in the relaxed micromorphic model: exploring finite-size metastructures for elastic wave control},
  author={Rizzi, Gianluca and d'Agostino, Marco Valerio and Neff, Patrizio and Madeo, Angela},
  journal={(arXiv:2105.00963) to appear in Mathematics and Mechanics of Solids},
  year={2021}
}

@article{aivaliotis_frequency-_2020,
	title = {Frequency- and angle-dependent scattering of a finite-sized meta-structure via the relaxed micromorphic model},
	volume = {90},
	issn = {0939-1533, 1432-0681},
	url = {http://link.springer.com/10.1007/s00419-019-01651-9},
	doi = {10.1007/s00419-019-01651-9},
	language = {en},
	number = {5},
	urldate = {2021-04-23},
	journal = {Archive of Applied Mechanics},
	author = {Aivaliotis, Alexios and Tallarico, Domenico and d’Agostino, Marco-Valerio and Daouadji, Ali and Neff, Patrizio and Madeo, Angela},
	month = may,
	year = {2020},
	pages = {1073--1096},
}

@article{neff_identification_2020,
	title = {Identification of {scale}-{independent} {material} {parameters} in the {relaxed} {micromorphic} {model} {through} {model}-{adapted} {first} {order} {homogenization}},
	volume = {139},
	issn = {0374-3535, 1573-2681},
	url = {http://link.springer.com/10.1007/s10659-019-09752-w},
	doi = {10.1007/s10659-019-09752-w},
	language = {en},
	number = {2},
	urldate = {2021-04-23},
	journal = {Journal of Elasticity},
	author = {Neff, Patrizio and Eidel, Bernhard and d’Agostino, Marco Valerio and Madeo, Angela},
	month = may,
	year = {2020},
	pages = {269--298},
}

@article{krushynska2017coupling,
  title={Coupling local resonance with Bragg band gaps in single-phase mechanical metamaterials},
  author={Krushynska, Anastasiia O and Miniaci, Marco and Bosia, Federico and Pugno, Nicola M},
  journal={Extreme Mechanics Letters},
  volume={12},
  pages={30--36},
  year={2017},
  publisher={Elsevier}
}

@article{willis_exact_2009,
	title = {Exact effective relations for dynamics of a laminated body},
	volume = {41},
	issn = {01676636},
	url = {https://linkinghub.elsevier.com/retrieve/pii/S0167663609000118},
	doi = {10.1016/j.mechmat.2009.01.010},
	language = {en},
	number = {4},
	urldate = {2021-04-23},
	journal = {Mechanics of Materials},
	author = {Willis, J. R.},
	month = apr,
	year = {2009},
	pages = {385--393},
}

@article{willis_effective_2011,
	title = {Effective constitutive relations for waves in composites and metamaterials},
	volume = {467},
	issn = {1364-5021, 1471-2946},
	url = {https://royalsocietypublishing.org/doi/10.1098/rspa.2010.0620},
	doi = {10.1098/rspa.2010.0620},
	language = {en},
	number = {2131},
	urldate = {2021-04-23},
	journal = {Proceedings of the Royal Society A: Mathematical, Physical and Engineering Sciences},
	author = {Willis, J. R.},
	month = jul,
	year = {2011},
	pages = {1865--1879},
}

@article{willis_construction_2012,
	title = {The construction of effective relations for waves in a composite},
	volume = {340},
	issn = {16310721},
	url = {https://linkinghub.elsevier.com/retrieve/pii/S1631072112000381},
	doi = {10.1016/j.crme.2012.02.001},
	language = {en},
	number = {4-5},
	urldate = {2021-04-23},
	journal = {Comptes Rendus Mécanique},
	author = {Willis, J. R.},
	month = apr,
	year = {2012},
	pages = {181--192},
}

@article{craster_high_frequency_2010,
	title = {High-frequency homogenization for periodic media},
	volume = {466},
	issn = {1364-5021, 1471-2946},
	url = {https://royalsocietypublishing.org/doi/10.1098/rspa.2009.0612},
	doi = {10.1098/rspa.2009.0612},
	language = {en},
	number = {2120},
	urldate = {2021-04-23},
	journal = {Proceedings of the Royal Society A: Mathematical, Physical and Engineering Sciences},
	author = {Craster, R. V. and Kaplunov, J. and Pichugin, A. V.},
	month = aug,
	year = {2010},
	pages = {2341--2362},
}

@article{nolde_high_2011,
	title = {High frequency homogenization for structural mechanics},
	volume = {59},
	number = {3},
	journal = {Journal of the Mechanics and Physics of Solids},
	author = {Nolde, E. and Craster, R. V. and Kaplunov, J.},
	year = {2011},
	pages = {651--671},
}

@article{sridhar_general_2018,
	title = {A general multiscale framework for the emergent effective elastodynamics of metamaterials},
	volume = {111},
	issn = {00225096},
	url = {https://linkinghub.elsevier.com/retrieve/pii/S0022509617306245},
	doi = {10.1016/j.jmps.2017.11.017},
	language = {en},
	urldate = {2021-04-23},
	journal = {Journal of the Mechanics and Physics of Solids},
	author = {Sridhar, A. and Kouznetsova, V.G. and Geers, M.G.D.},
	month = feb,
	year = {2018},
	pages = {414--433},
}

@article{pham_transient_2013,
	title = {Transient computational homogenization for heterogeneous materials under dynamic excitation},
	volume = {61},
	issn = {00225096},
	url = {https://linkinghub.elsevier.com/retrieve/pii/S0022509613001269},
	doi = {10.1016/j.jmps.2013.07.005},
	language = {en},
	number = {11},
	urldate = {2021-04-23},
	journal = {Journal of the Mechanics and Physics of Solids},
	author = {Pham, K. and Kouznetsova, V.G. and Geers, M.G.D.},
	month = nov,
	year = {2013},
	pages = {2125--2146},
}

@article{sridhar_homogenization_2016,
	title = {Homogenization of locally resonant acoustic metamaterials towards an emergent enriched continuum},
	volume = {57},
	issn = {0178-7675, 1432-0924},
	url = {http://link.springer.com/10.1007/s00466-015-1254-y},
	doi = {10.1007/s00466-015-1254-y},
	language = {en},
	number = {3},
	urldate = {2021-04-23},
	journal = {Computational Mechanics},
	author = {Sridhar, A. and Kouznetsova, V. G. and Geers, M. G. D.},
	month = mar,
	year = {2016},
	pages = {423--435},
}

@article{sridhar_frequency_2020,
	title = {Frequency domain boundary value problem analyses of acoustic metamaterials described by an emergent generalized continuum},
	volume = {65},
	issn = {0178-7675, 1432-0924},
	url = {http://link.springer.com/10.1007/s00466-019-01795-z},
	doi = {10.1007/s00466-019-01795-z},
	language = {en},
	number = {3},
	urldate = {2021-04-23},
	journal = {Computational Mechanics},
	author = {Sridhar, A. and Kouznetsova, V. G. and Geers, M. G. D.},
	month = mar,
	year = {2020},
	pages = {789--805},
}

@article{celli_bandgap_2019,
	title = {Bandgap widening by disorder in rainbow metamaterials},
	volume = {114},
	issn = {0003-6951, 1077-3118},
	url = {http://aip.scitation.org/doi/10.1063/1.5081916},
	doi = {10.1063/1.5081916},
	language = {en},
	number = {9},
	urldate = {2021-04-23},
	journal = {Applied Physics Letters},
	author = {Celli, Paolo and Yousefzadeh, Behrooz and Daraio, Chiara and Gonella, Stefano},
	month = mar,
	year = {2019},
	pages = {091903},
}

@article{bilal_architected_2018,
	title = {Architected {lattices} for {simultaneous} {broadband} {attenuation} of {airborne} {sound} and {mechanical} {vibrations} in {all} {directions}},
	volume = {10},
	issn = {2331-7019},
	url = {https://link.aps.org/doi/10.1103/PhysRevApplied.10.054060},
	doi = {10.1103/PhysRevApplied.10.054060},
	language = {en},
	number = {5},
	urldate = {2021-04-23},
	journal = {Physical Review Applied},
	author = {Bilal, Osama R. and Ballagi, David and Daraio, Chiara},
	month = nov,
	year = {2018},
	pages = {054060},
}

@article{liu_locally_2000,
  title={Locally resonant sonic materials},
  author={Liu, Zhengyou and Zhang, Xixiang and Mao, Yiwei and Zhu, YY and Yang, Zhiyu and Chan, Che Ting and Sheng, Ping},
  journal={Science},
  volume={289},
  number={5485},
  pages={1734--1736},
  year={2000},
  publisher={American Association for the Advancement of Science}
}

@article{wang_harnessing_2014,
	title = {Harnessing {buckling} to {design} {tunable} {locally} {resonant} {acoustic} {metamaterials}},
	volume = {113},
	issn = {0031-9007, 1079-7114},
	url = {https://link.aps.org/doi/10.1103/PhysRevLett.113.014301},
	doi = {10.1103/PhysRevLett.113.014301},
	language = {en},
	number = {1},
	urldate = {2021-04-23},
	journal = {Physical Review Letters},
	author = {Wang, Pai and Casadei, Filippo and Shan, Sicong and Weaver, James C. and Bertoldi, Katia},
	month = jul,
	year = {2014},
	pages = {014301},
}

@article{buckmann_mechanical_2015,
	title = {Mechanical cloak design by direct lattice transformation},
	volume = {112},
	issn = {0027-8424, 1091-6490},
	url = {http://www.pnas.org/lookup/doi/10.1073/pnas.1501240112},
	doi = {10.1073/pnas.1501240112},
	language = {en},
	number = {16},
	urldate = {2021-04-23},
	journal = {Proceedings of the National Academy of Sciences},
	author = {Bückmann, Tiemo and Kadic, Muamer and Schittny, Robert and Wegener, Martin},
	month = apr,
	year = {2015},
	pages = {4930--4934},
}

@article{misseroni_cymatics_2016,
	title = {Cymatics for the cloaking of flexural vibrations in a structured plate},
	volume = {6},
	issn = {2045-2322},
	url = {http://www.nature.com/articles/srep23929},
	doi = {10.1038/srep23929},
	language = {en},
	number = {1},
	urldate = {2021-04-23},
	journal = {Scientific Reports},
	author = {Misseroni, D. and Colquitt, D. J. and Movchan, A. B. and Movchan, N. V. and Jones, I. S.},
	month = apr,
	year = {2016},
	pages = {23929},
}

@article{cummer_controlling_2016,
	title = {Controlling sound with acoustic metamaterials},
	volume = {1},
	issn = {2058-8437},
	url = {http://www.nature.com/articles/natrevmats20161},
	doi = {10.1038/natrevmats.2016.1},
	language = {en},
	number = {3},
	urldate = {2021-04-23},
	journal = {Nature Reviews Materials},
	author = {Cummer, Steven A. and Christensen, Johan and Alù, Andrea},
	month = mar,
	year = {2016},
	pages = {16001},
}

@article{guenneau_acoustic_2007,
	title = {Acoustic metamaterials for sound focusing and confinement},
	volume = {9},
	issn = {1367-2630},
	url = {https://iopscience.iop.org/article/10.1088/1367-2630/9/11/399},
	doi = {10.1088/1367-2630/9/11/399},
	number = {11},
	urldate = {2021-04-23},
	journal = {New Journal of Physics},
	author = {Guenneau, Sébastien and Movchan, Alexander and Pétursson, Gunnar and Anantha Ramakrishna, S},
	month = nov,
	year = {2007},
	pages = {399--399},
}

@article{kaina_slow_2017,
	title = {Slow waves in locally resonant metamaterials line defect waveguides},
	volume = {7},
	issn = {2045-2322},
	url = {http://www.nature.com/articles/s41598-017-15403-8},
	doi = {10.1038/s41598-017-15403-8},
	language = {en},
	number = {1},
	urldate = {2021-04-23},
	journal = {Scientific Reports},
	author = {Kaina, Nadège and Causier, Alexandre and Bourlier, Yoan and Fink, Mathias and Berthelot, Thomas and Lerosey, Geoffroy},
	month = dec,
	year = {2017},
	pages = {15105},
}

@article{tallarico_edge_2017,
	title = {Edge {waves} and {localization} in {lattices} {containing} {tilted} {resonators}},
	volume = {4},
	issn = {2296-8016},
	url = {http://journal.frontiersin.org/article/10.3389/fmats.2017.00016/full},
	doi = {10.3389/fmats.2017.00016},
	urldate = {2021-04-23},
	journal = {Frontiers in Materials},
	author = {Tallarico, Domenico and Trevisan, Alessio and Movchan, Natalia V. and Movchan, Alexander B.},
	month = jun,
	year = {2017},
	pages = {16},
}

@article{willis_negative_2016,
	title = {Negative refraction in a laminate},
	volume = {97},
	issn = {00225096},
	url = {https://linkinghub.elsevier.com/retrieve/pii/S0022509615302623},
	doi = {10.1016/j.jmps.2015.11.004},
	language = {en},
	urldate = {2021-04-23},
	journal = {Journal of the Mechanics and Physics of Solids},
	author = {Willis, J. R.},
	month = dec,
	year = {2016},
	pages = {10--18},
}

@article{wen_effects_2011,
	title = {Effects of locally resonant modes on underwater sound absorption in viscoelastic materials},
	volume = {130},
	issn = {0001-4966},
	url = {http://asa.scitation.org/doi/10.1121/1.3621074},
	doi = {10.1121/1.3621074},
	language = {en},
	number = {3},
	urldate = {2021-04-23},
	journal = {The Journal of the Acoustical Society of America},
	author = {Wen, Jihong and Zhao, Honggang and Lv, Linmei and Yuan, Bo and Wang, Gang and Wen, Xisen},
	month = sep,
	year = {2011},
	pages = {1201--1208},
}

@article{mitchell_metaconcrete_2014,
	title = {Metaconcrete: designed aggregates to enhance dynamic performance},
	volume = {65},
	issn = {00225096},
	shorttitle = {Metaconcrete},
	url = {https://linkinghub.elsevier.com/retrieve/pii/S0022509614000131},
	doi = {10.1016/j.jmps.2014.01.003},
	language = {en},
	urldate = {2021-04-23},
	journal = {Journal of the Mechanics and Physics of Solids},
	author = {Mitchell, Stephanie J. and Pandolfi, Anna and Ortiz, Michael},
	month = apr,
	year = {2014},
	pages = {69--81},
}

@article{zhu_total-internal-reflection_2018,
	title = {Total-internal-reflection elastic metasurfaces: {Design} and application to structural vibration isolation},
	volume = {113},
	issn = {0003-6951, 1077-3118},
	shorttitle = {Total-internal-reflection elastic metasurfaces},
	url = {http://aip.scitation.org/doi/10.1063/1.5052538},
	doi = {10.1063/1.5052538},
	language = {en},
	number = {22},
	urldate = {2021-04-23},
	journal = {Applied Physics Letters},
	author = {Zhu, Hongfei and Walsh, Timothy F. and Semperlotti, Fabio},
	month = nov,
	year = {2018},
	pages = {221903},
}

@article{zhu_negative_2014,
	title = {Negative refraction of elastic waves at the deep-subwavelength scale in a single-phase metamaterial},
	volume = {5},
	issn = {2041-1723},
	url = {http://www.nature.com/articles/ncomms6510},
	doi = {10.1038/ncomms6510},
	language = {en},
	number = {1},
	urldate = {2021-04-23},
	journal = {Nature Communications},
	author = {Zhu, R. and Liu, X. N. and Hu, G. K. and Sun, C. T. and Huang, G. L.},
	month = dec,
	year = {2014},
	pages = {5510},
}

@article{kaina_negative_2015,
	title = {Negative refractive index and acoustic superlens from multiple scattering in single negative metamaterials},
	volume = {525},
	issn = {0028-0836, 1476-4687},
	url = {http://www.nature.com/articles/nature14678},
	doi = {10.1038/nature14678},
	language = {en},
	number = {7567},
	urldate = {2021-04-23},
	journal = {Nature},
	author = {Kaina, Nadège and Lemoult, Fabrice and Fink, Mathias and Lerosey, Geoffroy},
	month = sep,
	year = {2015},
	pages = {77--81},
}

@article{gonella_interplay_2009,
	title = {Interplay between phononic bandgaps and piezoelectric microstructures for energy harvesting},
	volume = {57},
	issn = {00225096},
	url = {https://linkinghub.elsevier.com/retrieve/pii/S0022509608001919},
	doi = {10.1016/j.jmps.2008.11.002},
	language = {en},
	number = {3},
	urldate = {2021-04-23},
	journal = {Journal of the Mechanics and Physics of Solids},
	author = {Gonella, Stefano and To, Albert C. and Liu, Wing Kam},
	month = mar,
	year = {2009},
	pages = {621--633},
}

@article{zhou_elastic_2008,
	title = {Elastic wave transparency of a solid sphere coated with metamaterials},
	volume = {77},
	issn = {1098-0121, 1550-235X},
	url = {https://link.aps.org/doi/10.1103/PhysRevB.77.024101},
	doi = {10.1103/PhysRevB.77.024101},
	language = {en},
	number = {2},
	urldate = {2021-04-23},
	journal = {Physical Review B},
	author = {Zhou, Xiaoming and Hu, Gengkai and Lu, Tianjian},
	month = jan,
	year = {2008},
	pages = {024101},
}

@article{zhang_asymmetric_2020,
	title = {An asymmetric elastic metamaterial model for elastic wave cloaking},
	volume = {135},
	issn = {00225096},
	url = {https://linkinghub.elsevier.com/retrieve/pii/S0022509619308762},
	doi = {10.1016/j.jmps.2019.103796},
	language = {en},
	urldate = {2021-04-23},
	journal = {Journal of the Mechanics and Physics of Solids},
	author = {Zhang, H.K. and Chen, Y. and Liu, X.N. and Hu, G.K.},
	month = feb,
	year = {2020},
	pages = {103796},
}

@article{chen_dispersive_2001,
	title = {A {dispersive} {model} for {wave} {propagation} in {periodic} {heterogeneous} {media} {based} on {homogenization} {with} {multiple} {spatial} and {temporal} {scales}},
	volume = {68},
	issn = {0021-8936, 1528-9036},
	url = {https://asmedigitalcollection.asme.org/appliedmechanics/article/68/2/153/458604/A-Dispersive-Model-for-Wave-Propagation-in},
	doi = {10.1115/1.1357165},
	language = {en},
	number = {2},
	urldate = {2021-04-23},
	journal = {Journal of Applied Mechanics},
	author = {Chen, W. and Fish, J.},
	month = mar,
	year = {2001},
	pages = {153--161},
}

@article{boutin_large_2014,
	title = {Large scale modulation of high frequency waves in periodic elastic composites},
	volume = {70},
	issn = {00225096},
	url = {https://linkinghub.elsevier.com/retrieve/pii/S0022509614001069},
	doi = {10.1016/j.jmps.2014.05.015},
	language = {en},
	urldate = {2021-04-23},
	journal = {Journal of the Mechanics and Physics of Solids},
	author = {Boutin, Claude and Rallu, Antoine and Hans, Stephane},
	month = oct,
	year = {2014},
	pages = {362--381},
}

@article{willis_variational_1981,
	title = {Variational principles for dynamic problems for inhomogeneous elastic media},
	volume = {3},
	issn = {01652125},
	url = {https://linkinghub.elsevier.com/retrieve/pii/0165212581900081},
	doi = {10.1016/0165-2125(81)90008-1},
	language = {en},
	number = {1},
	urldate = {2021-04-23},
	journal = {Wave Motion},
	author = {Willis, J. R.},
	month = jan,
	year = {1981},
	pages = {1--11},
}

@article{nassar_willis_2015,
	title = {Willis elastodynamic homogenization theory revisited for periodic media},
	volume = {77},
	issn = {00225096},
	url = {https://linkinghub.elsevier.com/retrieve/pii/S0022509614002579},
	doi = {10.1016/j.jmps.2014.12.011},
	language = {en},
	urldate = {2021-04-23},
	journal = {Journal of the Mechanics and Physics of Solids},
	author = {Nassar, H. and He, Q.-C. and Auffray, N.},
	month = apr,
	year = {2015},
	pages = {158--178},
}

@article{touboul2020effective,
  title={Effective resonant model and simulations in the time-domain of wave scattering from a periodic row of highly-contrasted inclusions},
  author={Touboul, Marie and Pham, Kim and Maurel, Agn{\`e}s and Marigo, Jean-Jacques and Lombard, Bruno and Bellis, C{\'e}dric},
  journal={Journal of Elasticity},
  volume={142},
  number={1},
  pages={53--82},
  year={2020},
  publisher={Springer}
}

@article{maurel2019multimodal,
  title={Multimodal method for the scattering by an array of plates connected to an elastic half-space},
  author={Maurel, Agn{\`e}s and Pham, Kim},
  journal={The Journal of the Acoustical Society of America},
  volume={146},
  number={6},
  pages={4402--4412},
  year={2019},
  publisher={Acoustical Society of America}
}

@article{fang2006ultrasonic,
  title={Ultrasonic metamaterials with negative modulus},
  author={Fang, Nicholas and Xi, Dongjuan and Xu, Jianyi and Ambati, Muralidhar and Srituravanich, Werayut and Sun, Cheng and Zhang, Xiang},
  journal={Nature materials},
  volume={5},
  number={6},
  pages={452--456},
  year={2006},
  publisher={Nature Publishing Group}
}

@article{dubois2013flat,
  title={Flat lens for pulse focusing of elastic waves in thin plates},
  author={Dubois, Marc and Farhat, Mohamed and Bossy, Emmanuel and Enoch, Stefan and Guenneau, S{\'e}bastien and Sebbah, Patrick},
  journal={Applied Physics Letters},
  volume={103},
  number={7},
  pages={071915},
  year={2013},
  publisher={American Institute of Physics}
}

@article{jin2016gradient,
  title={Gradient index devices for the full control of elastic waves in plates},
  author={Jin, Yabin and Torrent, Daniel and Pennec, Yan and Pan, Yongdong and Djafari-Rouhani, Bahram},
  journal={Scientific reports},
  volume={6},
  number={1},
  pages={1--8},
  year={2016},
  publisher={Nature Publishing Group}
}

@article{colombi2016seismic,
  title={A seismic metamaterial: The resonant metawedge},
  author={Colombi, Andrea and Colquitt, Daniel and Roux, Philippe and Guenneau, Sebastien and Craster, Richard V},
  journal={Scientific reports},
  volume={6},
  number={1},
  pages={1--6},
  year={2016},
  publisher={Nature Publishing Group}
}

@article{du2017elastic,
  title={Elastic metamaterial-based seismic shield for both Lamb and surface waves},
  author={Du, Qiujiao and Zeng, Yi and Huang, Guoliang and Yang, Hongwu},
  journal={AIP Advances},
  volume={7},
  number={7},
  pages={075015},
  year={2017},
  publisher={AIP Publishing LLC}
}

@article{miniaci2016large,
  title={Large scale mechanical metamaterials as seismic shields},
  author={Miniaci, Marco and Krushynska, Anastasiia and Bosia, Federico and Pugno, Nicola M},
  journal={New Journal of Physics},
  volume={18},
  number={8},
  pages={083041},
  year={2016},
  publisher={IOP Publishing}
}

@article{achaoui2016seismic,
  title={Seismic waves damping with arrays of inertial resonators},
  author={Achaoui, Younes and Ungureanu, Bogdan and Enoch, Stefan and Br{\^u}l{\'e}, St{\'e}phane and Guenneau, S{\'e}bastien},
  journal={Extreme Mechanics Letters},
  volume={8},
  pages={30--37},
  year={2016},
  publisher={Elsevier}
}

@article{neff2014unifying,
  title={A unifying perspective: the relaxed linear micromorphic continuum},
  author={Neff, Patrizio and Ghiba, Ionel-Dumitrel and Madeo, Angela and Placidi, Luca and Rosi, Giuseppe},
  journal={Continuum Mechanics and Thermodynamics},
  volume={26},
  number={5},
  pages={639--681},
  year={2014},
  publisher={Springer}
}
\endgroup


\end{document}